\documentclass[twoside,11pt]{article}
\usepackage{theapa}
\usepackage{jair}
\usepackage{rawfonts}
\usepackage{url}
\usepackage{graphicx}
\usepackage{multirow}
\usepackage{makecell}
\usepackage{tabularray}
\usepackage{booktabs}
\usepackage{longtable}
\usepackage{xcolor}
\usepackage{caption}
\usepackage{subcaption}
\usepackage{microtype}
\usepackage{float}

\jairheading{X}{XXXX}{X-XX}{5/15}{X/XX}
\ShortHeadings{Trust in AI-DSS and Validation of MAST}
{Chiou et al.}
\firstpageno{1}

\begin{document}

\title{Evaluating Trustworthiness of AI-Enabled Decision Support Systems: 
Validation of the Multisource AI Scorecard Table (MAST)}

\author{\name Pouria Salehi \email psalehi@asu.edu \\
        \name Yang Ba \email yangba@asu.edu \\
        \name Nayoung Kim \email nkim48@asu.edu \\
        \name Ahmadreza Mosallanezhad \email amosalla@asu.edu \\
        \name Anna Pan \email arolso10@asu.edu\\
        \name Myke C. Cohen \email myke.cohen@asu.edu \\
       \name Yixuan Wang \email ywan1290@asu.edu \\
       \name Jieqiong Zhao \email jzhao153@asu.edu\\
       \name Shawaiz Bhatti \email sabhatt1@asu.edu \\
       \name Michelle V. Mancenido \email mvmancenido@asu.edu \\
       \name Erin K. Chiou \email echiou@asu.edu \\ 
       \addr Arizona State University, USA\\
       \AND
       \name Erik Blasch \email erik.blasch.1@us.af.mil  \\
       \addr Air Force Office of Scientific Research,
       USA
       \AND
       \name James Sung \email james.sung@hq.dhs.gov\\ 
       \addr DHS Office of Intelligence and Analysis,
       USA}


\maketitle

\begin{abstract}
The Multisource AI Scorecard Table (MAST) is a checklist tool based on analytic tradecraft standards to inform the design and evaluation of trustworthy AI systems. In this study, we evaluate whether MAST is associated with people’s trust perceptions in AI-enabled decision support systems (AI-DSSs).
Evaluating trust in AI-DSSs poses challenges to researchers and practitioners. These challenges include identifying the components, capabilities, and potential of these systems, many of which are based on the complex deep learning algorithms that drive DSS performance and preclude complete manual inspection. 
We developed two interactive, AI-DSS test environments using the MAST criteria. One emulated an identity verification task in security screening, and another emulated a text summarization system to aid in an investigative reporting task. Each test environment had one version designed to match low-MAST ratings, and another designed to match high-MAST ratings, with the hypothesis that MAST ratings would be positively related to the trust ratings of these systems. A total of 177 subject matter experts were recruited to interact with and evaluate these systems. 
Results generally show higher MAST ratings for the high-MAST conditions compared to the low-MAST groups, and that measures of trust perception are highly correlated with the MAST ratings.
We conclude that MAST can be a useful tool for designing and evaluating systems that will engender high trust perceptions, including AI-DSS that may be used to support visual screening and text summarization tasks. However, higher MAST ratings may not translate to higher joint performance.  

\end{abstract}

\section{Introduction}
\label{Introduction}
Decision-making is increasingly dependent on artificial intelligence (AI) in many high-stakes domains \cite{phillips2012ai}. In healthcare, for example, AI systems are now used to assess breast lesions and cancerous growths, detect and diagnose arrhythmia from ECG readings, and generate radiation treatment plans for cancer patients with minimal human inputs \cite{zhu20222021}. AI-enabled decision support systems (AI-DSSs) like these allow strained institutions and agencies to meet growing demands for services that depend on limited human resources \cite{knop2022human}. However, concerns about the trustworthiness of AI-DSSs are rising in safety-critical areas such as airport security, medical diagnostics, and national defense, where poor performance could result in catastrophic outcomes \cite{cooke2007stories}. To safeguard ethical and value-based considerations for the general public, AI systems in these domains are typically designed as human-in-the-loop systems. 

Human-in-the-loop designs often entail human supervision over imperfect AI decision-making processes to ensure that decisions align with ethical standards and the best interests of society \cite{parasuraman2008humans}. However, trust can be a factor in human-in-the-loop decisions because trust impacts the willingness to work with the system. People's trust is closely related to their confidence in the a system's capabilities, including the ability to make decisions consistent with their values \cite{lee2004trust}. In safety-critical and time-constrained task environments, trust in AI can be crucial when people need to rely on AI recommendations or actions without consistent monitoring or second-guessing. Thus, trustworthy human-in-the-loop AI-DSSs must balance human supervisors' abilities to safeguard ethical impact, while also allowing them to generally use the AI system as a reliable tool. Both modes are essential to ensure the responsible and effective use of AI technology in safety-critical scenarios.


Existing design frameworks and best practice guidelines \cite<e.g.,>{de2020towards,schaefer2016meta} for human-AI systems are generally broad-stroked in their recommendations. The challenge of translating those recommendations into operationalized features of AI technologies has tested the overall practicality and impact of these frameworks on both developing and built systems. More specific guidance that can be quickly operationalized to effectively guide design or evaluation of AI-enabled systems, with minimal design-test-evaluation cycles required to get to effectiveness, remains an ongoing pursuit for both researchers and practitioners. 

The Multisource AI Scorecard Table (MAST; \citeR{sung2019national,blasch2021multisource}) is a checklist tool that was developed in part to address this gap, and aid the design and evaluation of transparent and trustworthy AI systems. MAST was developed based on the Intelligence Community Directive (ICD) 203, which describes a set of nine tradecraft standards adopted by the Intelligence Community to evaluate the quality of human intelligence reporting \cite{odniIntelligenceCommunityDirective2015}. These criteria include: sourcing, uncertainty, distinguishing, analysis of alternatives, customer relevance, logical argumentation, consistency, accuracy, and visualization. Following ICD 203, MAST also addresses data transformation, data aggregation and labeling,  data display, and decision relevance; encompassing various phases of the data life cycle contributing to AI systems from data collection and processing to mode training and output, and from verification and validation to operation and monitoring \cite{blasch2019artificial}. Specifically, data transformation includes sourcing, uncertainty, distinguishing, and analysis of alternatives, relating to data collection and processing. Data aggregation and labeling consist of logical argumentation, consistency, and accuracy and involve the model training process. Lastly, data display and decision relevance contain analysis of alternatives, customer relevance, and visualization, regarding verification and validation. The idea behind MAST is that considering these nine criteria in the design of AI systems should result in more transparent and trustworthy outputs, which in turn will make it more likely to be used and result in more effective joint human-automation performance. Although previous work has demonstrated the utility of MAST through case studies \cite{sung2019national,blasch2021multisource}, no empirical study dedicated to validating this tool has been published as of the time of writing.

This paper pursues this objective by using MAST to inform both the design and subsequent evaluation of two distinct AI-DSSs: an image processing system called Facewise, and a language processing and data visualization system called the REporting Assistant for Defense and Intelligence Tasks (READIT). The goal of this study was to investigate (1) whether the design and evaluation of AI-enabled systems based on the MAST criteria can predict trust perceptions, and (2) whether MAST can also be used to evaluate the trustworthiness of AI-DSS in safety critical task environments more broadly (i.e., not just AI used for intelligence or reconnaissance tasks). 

Our study provides insights into the effectiveness of using MAST as a tool for designing and evaluating AI-enabled systems, and connects these insights with current trust scholarship. Our results suggest that incorporating the nine MAST criteria into the design of AI systems can improve trust perceptions of these systems. Additionally, MAST appears to be a useful tool for improving trust perceptions of AI-enabled systems beyond those designed for intelligence tasks. This paper also identifies potential limitations of MAST and areas for future research, including our finding that high MAST ratings do not necessarily lead to high joint performance. This finding is consistent with previous work suggesting that high trust ratings do not necessarily lead to improvements in joint system performance. Overall, this study highlights the ongoing challenge of operationalizing generalizable criteria in ways that can improve joint human-automation system performance, and the challenge of leveraging trust concepts in designing effective human-AI systems. At the same time, this study supports MAST as a promising tool for achieving AI design that is connected to practitioner norms, provides a mechanism for documenting relevant transparency information, and can result in high trust perceptions for systems intended for use in safety critical task environments.

\section{Background}
\label{Background}

The role of people as the final arbiters over imperfect automation has a long history \cite{Sheridan_1975,Bainbridge_1983}. In the supervisory control structures that govern most human-AI systems, people are tasked with assessing and, if necessary, intervening in AI outputs. However, many DSSs are designed for task environments in which people rarely have the cognitive and physical resources to sufficiently understand, assess, and intervene with every recommendation \cite{mcguirl2006supporting}. This is especially true in safety-critical systems, in which people may be expected to attend to every outcome produced by imperfect AI-DSSs. 

Limitations in human decision-making amid imperfect AI-DSSs have resulted in novel types of problematic outcomes, some of which have been catastrophic. For example, people tend to overly rely on decisions recommended by automation or AI, even when there are clear indications that the recommendation may be wrong (e.g., automation bias; \citeR{skitka1999does}). An infamous case is from the Iraq war, in which the Patriot missile system’s DSS erroneously identified allied fighter jets as enemy aircraft. Operators of the missile system approved the DSS-recommended decision to attack the aircraft, causing the fratricide of American and British pilots \cite{cummings2006automation}. More recently, a series of wrongful arrests in the United States has been traced to law enforcement reliance on facial recognition technologies that have considerable racial and gender biases \cite<e.g.,>{hill2020wrongfully,hill2023thousands}. However, upon recognizing errors in AI recommendations, there is also a tendency for people to reject future AI recommendations (e.g., automation aversion; \citeR{dietvorst2015algorithm}), especially by experts in the decision-making domain \cite{snow2021satisficing}.

People's tendency to overuse, misuse, or disuse DSS has long been linked to poorly calibrated perceptions of the DSS's trustworthiness with respect to its actual reliability \cite{parasuramanHumansAutomationUse1997}. As such, methodological frameworks, policy guidelines, and other tools for designing and evaluating DSS trustworthiness have proliferated alongside advancements in AI-DSS capabilities. These include but are not limited to, the Microsoft UX Design Principle \cite{microsoft1995windows}, NISTIR 8330 by National Institute of Standards and Technology \cite{stantonjensen2021}, AI Fairness 360 Toolkit by IBM \cite[and others]{bellamy2019ai}, IEEE Global Initiative on Ethics of Autonomous and Intelligent Systems \cite{chatila2019ieee}, UXPA Guidelines for Trustworthy User Experiences \cite{kriskovic2017dark}, or Ethical OS Toolkit \cite{lilley2020using}. Although these tools do not all explicitly focus on the concept of trust and trustworthiness, they share an underlying motivation that the design, development, and evaluation of AI systems that impact people and organizations require attention to human factors. 

Despite the existence of many frameworks and tools to guide the design of trustworthy AI and other software systems, designing for trust and evaluating trustworthiness with precision remains a challenge. There is a wide translation gap between theory and practice, partly because trust is an abstract construct with myriad closely related concepts. For example, designing trustworthy systems also often involves designing for transparency, individual differences, workload, situation awareness, and attending to other possible factors like etiquette and anthropomorphism \cite{hoff2015trust,parasuraman2004trust}. Another challenge to effectively designing trustworthy AI is that the various expert communities in different domains may define trust differently. These differences in definitions can be attributed to what each community values most and therefore, designing for trustworthy AI means something different for every community. For example, the intelligence community might value high-quality data as a foundation for high-quality analysis. For the transportation security community, it might value high-quality decisions made at the front lines that could affect traveler safety, more so than data integrity. 


To address this gap between concept and practice of designing and evaluating the trustworthiness of AI systems, the Multisource AI Scorecard Table (MAST; \citeR{sung2019national,blasch2021multisource}) was developed by the AI Team of the 2019 Public-Private Analytic Exchange Program, supported by the Office of the Director of National Intelligence and Department of Homeland Security. MAST describes nine criteria derived from analytic tradecraft standards ICD 203 to assess the trustworthiness of intelligence reporting, and additionally includes a four-level quantitative breakdown for each criterion. The idea is that MAST could serve as an easy-to-use checklist for designing trustworthy AI-enabled systems, and for evaluating trustworthiness after system development. Although the principles behind MAST would seem more suitable for intelligence tasks, given its focus on information quality and integrity, it is possible that these criteria may also be applied to other AI systems used for general information-processing and other human decision-making tasks. For example, AI-enabled systems in computer vision, natural language processing, and medical diagnostic tasks may all be rated according to MAST criteria, including rating the system’s sourcing (e.g., credibility of training data), or its ability to describe and propose alternative recommendations. Medical professionals and their patients may be more willing to trust an AI-derived diagnosis and treatment plan if the system was developed to include the MAST criteria of uncertainty, analysis of alternatives, and customer relevance.

It should be noted that several instruments have been developed to measure trust in general automation, including instances of AI-enabled automation \cite{Alsaid_Li_Chiou_Lee_2023,Kohn_deVisser_Wiese_Lee_Shaw_2021}. 
Many of these instruments have been widely adopted, others have been independently validated. However, these instruments were mainly designed for research or technology evaluation purposes, rather than for technology development or operational settings. Therefore, although these instruments could be considered relatively robust when used appropriately, they suffer from similar limitations as the design guidelines and frameworks for evaluating system trustworthiness described previously. There remain wide translation gaps, subject to highly variable interpretation from principles to practice, given the hundreds of under-specified conditions and decisions that system designers and other practitioners face. For example, underlying many of these instruments is a nuanced presumption that assessing domain experts' trust in a particular technology, after they have experienced using the technology, could be some indication of the technology's trustworthiness. This presumed connection between trust and trustworthiness is then flattened in some practitioner circles, in which they will equate high trust perceptions with technology trustworthiness, despite most trust experts being careful not to conflate the two. 

To situate the MAST tool in the context of current trust scholarship, our primary objective is to assess the construct validity of MAST relative to human trust. Construct validity is the degree to which an instrument measures the construct it was designed to measure \cite{cronbach1955construct}. Approaches for evaluating construct validity include multivariate analytical tools, such as factor analysis \cite{raykov2008introduction,tabachnick2013using}, principal components analysis (PCA; \citeR{bandalos2018measurement}), and structural equation modeling \cite{kline2015principles}. The goal of using multivariate analysis in construct validation is to capture, explain, and measure the amount of variation among items for a construct and to associate these with previously validated constructs \cite{chancey2017trust,jian2000foundations}. This study aimed to validate MAST as an instrument for assessing trust by investigating how MAST items are associated with validated trust questionnaires.
\section{General Method}
\label{General Method}

\subsection{Testbeds}
To validate MAST in different contexts, we used the MAST criteria as a framework, in conjunction with our expertise and knowledge on state-of-the-art AI algorithms, to design two AI-DSS testbeds: one for identity-verification in a security screening task (Facewise) and another for a text analysis task to support intelligence reporting (READIT). 
Facewise is an emulated 1-to-1 identity verification system built on a pre-trained convolutional neural network with extra fine-tuning on face recognition tasks with Cross-entropy loss. Facewise compares two face images, an identification photo and a live or encounter photo, and provides a recommendation on whether the two images are of the same identity (match) or of different identities (mismatch). These types of face-matching decision support systems powered by AI are becoming prevalent in airport security checkpoints, i.e., CAT-C or CAT2 \cite{lim2021privacy}. 

READIT is an emulated natural language processing system that was designed to compile, summarize, and categorize documents of limited length (news articles, reports, microblogs) to expedite intelligence gathering and reporting. READIT first uses BERT \cite{devlin-etal-2019-bert} to generate outputs, after which we manually improved on the model outputs to enhance the usefulness and usability of the tool. 

The case scenario developed for READIT was to assess MAST within the text summarization contexts that MAST was originally designed and evaluated for \cite{blasch2021multisource}. The identity verification scenario was selected to test the validity of the MAST criteria using a different type of AI capability, in a  different type of task environment, while staying within a national security context subject to low risk tolerance. Our case scenario and AI-DSS testbeds were designed and developed based on information gathered from field visits to operational security screening environments, and bi-monthly consultations with operational stakeholders (i.e., national security researchers, practitioners, and analysts).

Both Facewise and READIT were developed using cloud-based services consisting of client-server model for user-AI interaction. In the Facewise system, we leveraged Amazon Web Services (AWS) and Google Cloud Platform (GCP) for efficient use of storage and resources. We built the client part of the platform with HTML5 and JavaScript. We collected the responses from participants on the client’s side and sent them to the GCP through Python3 and Flask library to save them in the database. Similarly, the READIT system consisted of a JavaScript based client that enables the participant-AI interaction, and the server was built using Python3 and Flask library, hosted on GCP. Data visualizations on READIT were created to aid in better understanding of the dataset. The visualizations were implemented using D3.js, which is a popular open-source JavaScript library for creating custom interactive data visualizations. While participants were conducting the task, we logged system activities (e.g., button clicks, and relevant changes to the system state) to assess performance. The implementation code for READIT and Facewise is available at: https://github.com/nayoungkim94/PADTHAI-MM.


\subsection{Constructs and Measures}
For both DSS platforms, system features were manipulated to comprise two versions (High-MAST and Low-MAST) with eight outcome variables of interest: MAST criteria ratings; perceptions of risk, benefit, trust, credibility; task performance; self-reported engagement and usability. These constructs and measures are defined in more detail below. 

\paragraph{Versions of the DSS: High-MAST and Low-MAST.}
System features refer to the available features that a DSS can provide its operators. Based on the MAST criteria, two levels of features for each platform were created: High-MAST and Low-MAST. High-MAST features were designed to score high ratings on each of the MAST criteria, resulting in a set of rich features that was supposed to be helpful to excel in the task. On the other hand, Low-MAST features were designed to score low ratings on each of the MAST criteria with a minimum set of necessary features included to be able to complete a task. In summary, the High-MAST versions could be described as providing more information about the DSS's performance given the task context, and the Low-MAST versions were designed to operate more like black-box systems. However, both High- and Low-MAST versions were designed to be as equal as possible in terms of engagement and usability. Appendices A and B delineate the MAST criteria and detailed feature descriptions for Facewise and READIT, respectively. More information about our development process and design decisions of our DSS testbeds are not the focus of this paper, but will be reported in detail in a forthcoming paper. 

\paragraph{Variables of interest: MAST criteria, risk, benefit, trust, credibility, performance, engagement, and usability.}
Each DSS was evaluated based on descriptions of the MAST criteria and using a Likert-like scale of 1 to 4, with 1 being poor and 4 being excellent. Each MAST criterion was shown in a question format and accompanied by a corresponding feature description of the DSS. The MAST-total score was created by adding up these 9 criteria with a range of 9 to 36. Participant perception of risk and benefit was measured through two items derived from \cite{weber2002domain}. Risk was included due to the well-known relationship between trust and risk \cite{lee2004trust} and perceived benefit was included to check whether participants felt that using the DSS was beneficial for the task they were asked to complete. To measure trust, we used two commonly-used questionnaires. One is a previously validated, 12-item instrument known to measure general trust perceptions of automation \cite{jian2000foundations,spain2008towards}. The second 15-item instrument measures trust by querying about specific types of information known to affect trust -- purpose, process, and performance \cite{chancey2017trust}. Because the MAST criteria largely focus on the presentation, availability, types, and quality of information presented by the AI system, we also included a measurement for message credibility (i.e., excluding source credibility), adopting a 3-item survey \cite{appelman2016measuring}. Appendix C presents the scale, example items, number of items for each variable, and their Cronbach’s alpha. 

We measured performance through two variables, average task completion time and a scenario-specific performance metric. For Facewise, our scenario-specific performance metric focused on average accuracy on the identity verification task that took roughly 30 minutes to complete. The off-the-shelf performance of the algorithm used had an accuracy rate of 95\% across test data during model training \cite{8248937}. However, with the database used in this experiment, its performance was around 60\% for the difficult cases, and greater than 95\% for the easy cases. Participants were not given this information about the algorithm's performance in advance nor were they informed about the potential difficulty levels of the cases in advance, but they were alerted to the fact that part of their task was to ensure that the correct decision was made with an algorithm that was potentially fallible. More information about the algorithm's performance could be accessed by participants only from the additional information option provided as part of the High-MAST interface. 

For READIT, our scenario-specific performance metric was based on similarity to the ground truth adopted from the 2011 IEEE Visual Analytics Science and Technology (VAST) Challenge \cite{IEEEVAST2011MC3}. We asked participants to determine the cause of a terrorist threat in a fictitious city named ``Vastopolis" by using the READIT system to navigate the information in the available text documents (obtained from the VAST Challenge) and complete a 250-word report within roughly 60 minutes. We had four researchers independently rate participants' analytical reports in comparison to a ground truth description, and we averaged their ratings which largely converged to determine their performance score in the investigative reporting task. Other than task performance, the remaining dependent variables and covariates were identical for Facewise and READIT.  

To ensure that our implementation of different system features across the two testbeds would not cause major differences in perceived system usability and task engagement, and subsequently affect trust, risk, and benefit perceptions, we measured participants' perceived usability and engagement in the interactive task. Usability was assessed with a widely-used 10-item questionnaire known as the System Usability Scale \cite{brooke2020sus} and engagement was assessed with a 17-item questionnaire \cite{schaufeli2002measurement}. Appendix C reports more details of these measures. In addition, because study participants were experts (professionally trained in) face matching and intelligence analysis task domains, we manipulated task difficulty to ensure sufficient task engagement. For Facewise, we did this by hand-selecting 80 pairs of face images with known ground truth, with 40 of those image pairs representing easy tasks and the remaining 40 image pairs representing difficult tasks, presented in randomized order. Difficulty was defined from the perspective of the human operator, with difficult image pairs largely selected from a publicly available sibling database \cite{BMVC2015_41}, and validated in a pilot study with a general population sample that was on average more likely to get the difficult pairs wrong. For READIT, we included ``red herring'' documents from the VAST Challenge dataset \cite{IEEEVAST2011MC3} to encourage participant engagement in the task; several of these documents were related and collectively formed narratives that presented several plausible causes for the terrorist threat scenario. These ``red herring'' documents would then ideally cause sufficiently engaged participants to consider several highly plausible conclusions for their final report.

\subsection{Procedure}
Figure 1 illustrates the general procedure of this study for Facewise and READIT. We first created a virtual hub using the web-based software platform Qualtrics \cite{Qualtrics} for participants to access and complete the study. After random assignment to one of the two conditions (High-MAST or Low-MAST) by the researchers, participants were asked to input their given participant IDs on the first page of Qualtrics. Next, informed consent approved by our Institutional Review Board (IRB) was obtained. Then, participants were given a description of their task scenarios. To facilitate engagement and a sense of risk in the study scenario, in all conditions participants were told that they were being tasked to complete an important assignment, and that a previous agent assigned to their post failed in their respective tasks and was put on probation and subsequently demoted. After reading the task descriptions, participants then received a short training on their respective DSSs by watching a recorded video demonstration of the interface and features, and responding correctly to quiz questions about the video. All DSS versions were presented as technology aids that exist to supplement the participant's own abilities. Afterward, participants performed the study task using their respective DSS. Lastly, participants were asked to evaluate the system and their experience by responding to questionnaires including the MAST criteria, risk, benefit, trust, credibility, engagement, and usability. Given our targeted population of subject matter experts in national security, limited optional demographic information was collected to assess the representativeness of our sample population.

\begin{figure*}[t!]
    \centering
    \includegraphics[scale=0.43]{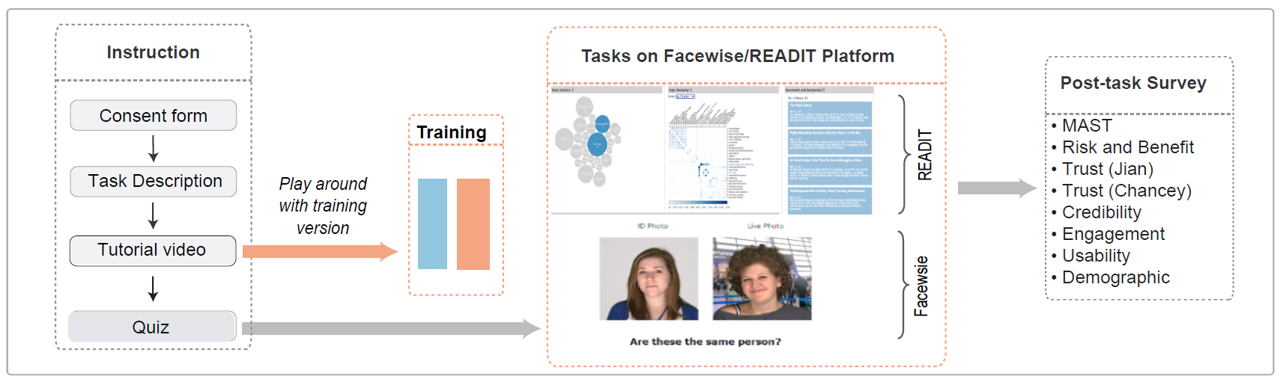}
    \caption{Study Procedure for Facewise and READIT.}
    \label{fig:fig1}
\end{figure*}

\subsection{Data Analysis}
Data analysis was accomplished in JMP \cite{JMP} and R using “dplyr” \cite{wickhamr}, “psych” \cite{revelle2018procedures}, “Rmisc” \cite{R-Rmisc}, and “compareGroups” \cite{subirana2014building}. Figures were created using “ggmap” \cite{kahle2013ggmap} and \cite{R-gridExtra}. To confirm associations between the MAST criteria and measures of trust, credibility, and other validated metrics, we performed the analysis in three steps. First, we explored if there are differences between the Low-MAST and High-MAST groups with respect to the dependent variables identified above. Secondly, linear associations between the individual metrics and the averaged MAST rating were separately established using simple linear regression. Multivariate analysis via principal components analysis (PCA) was then performed on the perceptual metrics for dimension reduction. Finally, we regressed the MAST ratings with the principal component (PC) scores. PCA was employed to find coherent and appropriate structures in the perceptual metrics within the first few principal components \cite{bandalos2018measurement}. To compare the different levels of Facewise and READIT, Analysis of Variance (ANOVA) was used. In addition, linear regression was used to further investigate the strength and directionality between MAST and other survey measures. 
\section{Experiment 1: Facewise}
\label{Experiment 1: Facewise}

Participants in the Facewise experiment were told that they were airport security officers tasked to screen passengers by checking their identification materials with the assistance of Facewise, and they had roughly 30 minutes to complete a series of identification verification tasks which is roughly the length of an officer's shift in the document checker position \cite{Greene_Kudrick_Muse_2014}. Figure 2 outlines the similarities and differences between the two levels of Facewise: High-MAST and Low-MAST. For both levels, Page 1 asks for an initial judgment of human operators. We adopted this structure based on previous work, which we found would increase accuracy \cite{salehi2021decision}. In Page 1 (Figure 2), the left image with an off-white background presents the ID photo and the right image with an airport background provides a “live” photo, supposedly taken at the airport. For both levels, these images were cropped and zoomed in for Page 2, which is where most of the differences between High-MAST and Low-MAST appear. Three red dotted lines highlight these differences including the Crossmark/Checkmarks, AI confidence, and a ``View Details'' button. For more details regarding the AI-DSS features and how they map to each of the MAST criteria, please refer to Appendix A. 

\begin{figure*}[t!]
    \centering
    \begin{subfigure}[t]{0.9\textwidth}
         \centering
         \includegraphics[width=\textwidth]{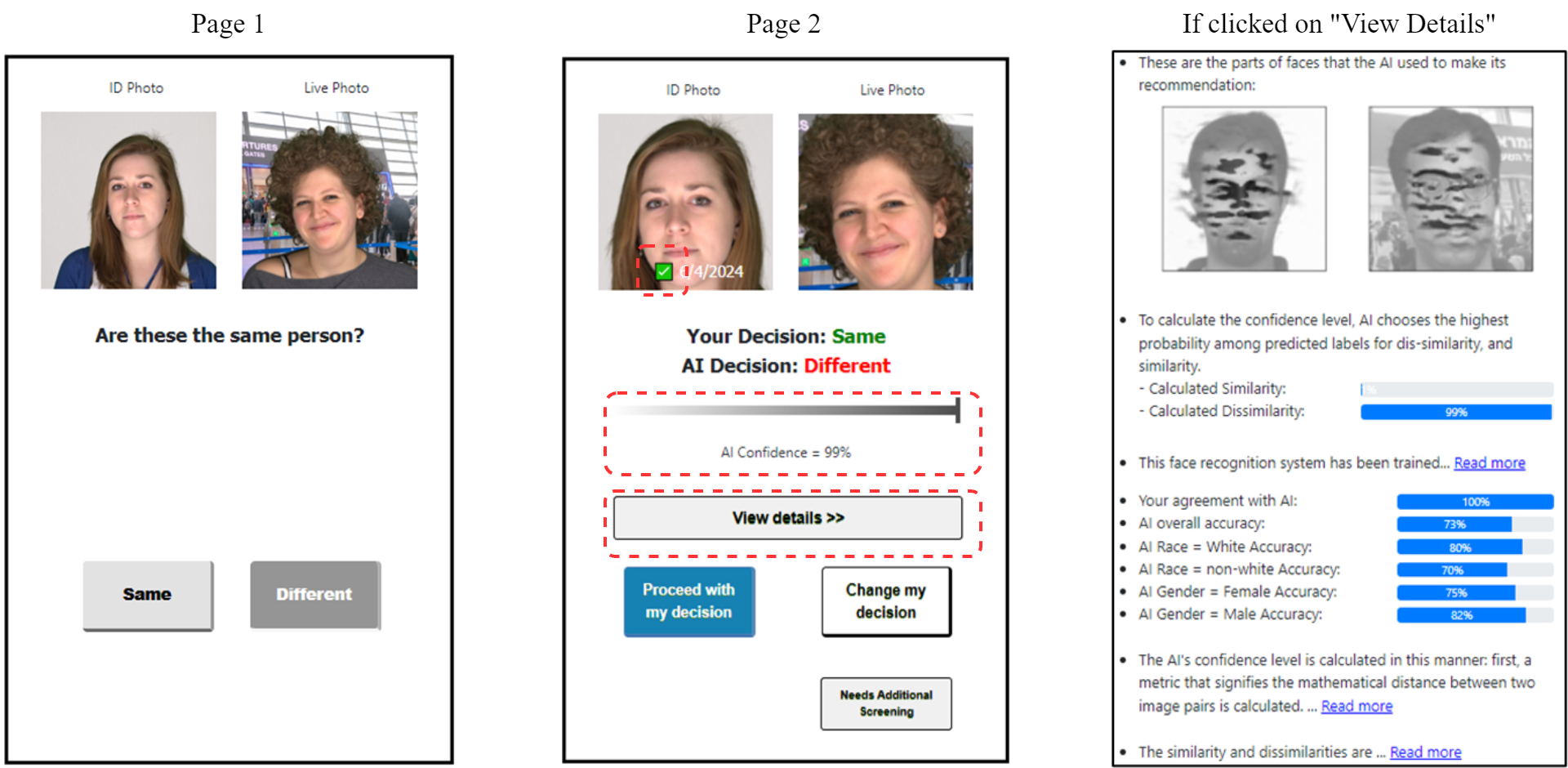}
         \caption{Facewise High-MAST}
         \label{fig:2a}
    \end{subfigure}

    \begin{subfigure}[t]{0.6\textwidth}
        \centering
        \includegraphics[width=\textwidth]{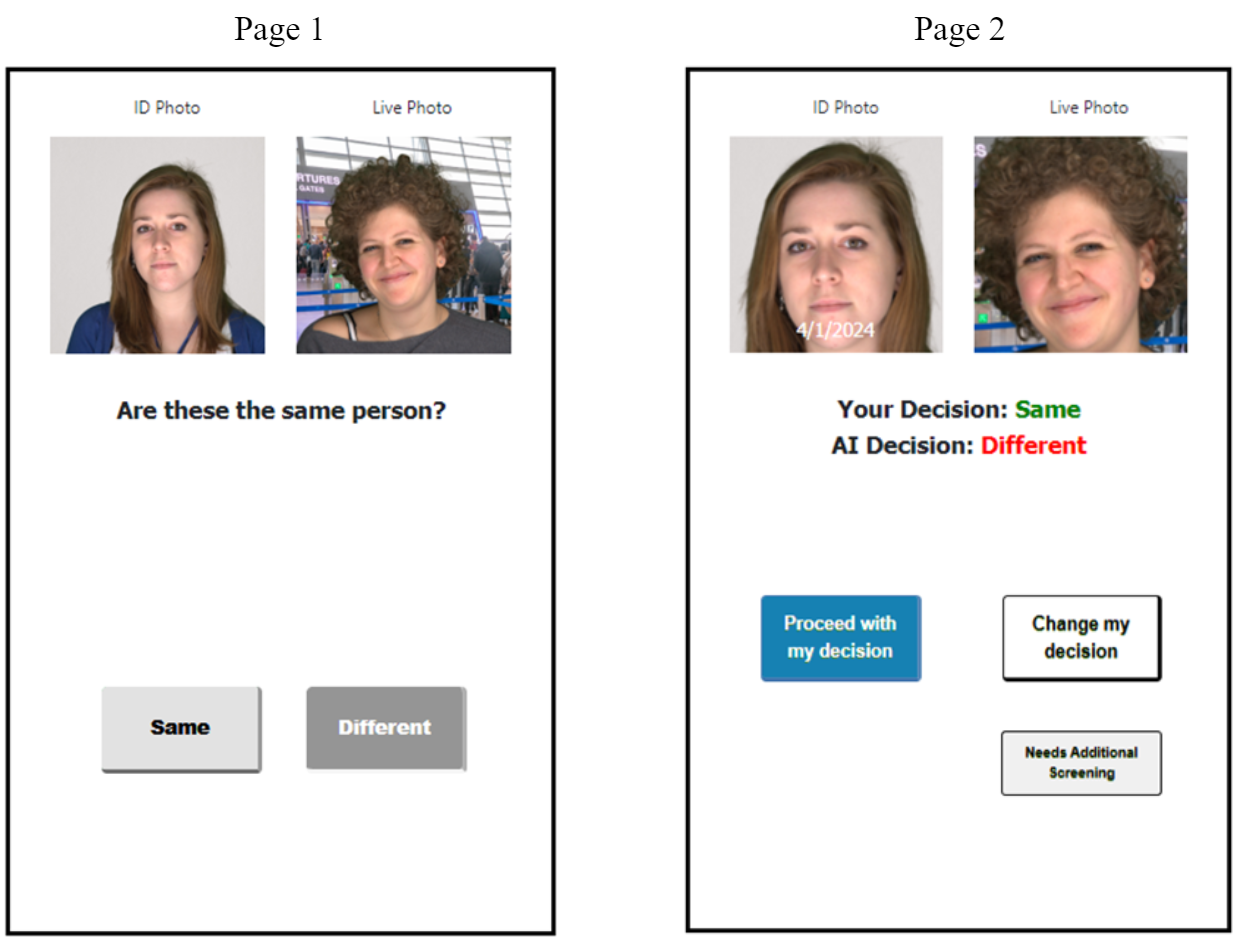}
        \caption{Facewise Low-MAST}
        \label{fig:2b}
    \end{subfigure}

    \caption{Facewise High-MAST (top) and Facewise Low-MAST (down). Red dashes highlight the differences between Low and High platforms. Compared to the Facewise Low-MAST, High-MAST version has more interactive features including the ID expiration check, AI confidence level, and ``View Details'' page.}
    \label{fig:fig2}
\end{figure*}

\subsection{Participants}
A total of 152 subject matter experts, U.S. Transportation Security Officers (TSOs), were recruited from three major U.S. airports in Arizona, Nevada, and California, split across 11 days of data collection at the participating airports. Six participants were removed due to very high response time and low accuracy, resulting in 73 participants each for the High-MAST and Low-MAST conditions. On average, participants spent 76 minutes to complete the entire study, including onboarding and responses to questionnaire items. Because participants were federal employees, we were not permitted to provide compensation despite their participation being voluntary and outside of their regular duties. Therefore, light refreshments were provided to appreciate their participation in the study. Because all participants were volunteers and not required to participate, we assumed they were sufficiently motivated by our study objectives to complete this study to the best of their ability. Table 5 in Appendix F reports the available participant demographics across the Facewise conditions. Race, ethnicity, and gender items were not collected or reported due to expressed concerns by some of our collaborative partners, given our limited population of subject matter experts, and the sensitive nature of this information.

\subsection{Results and Discussion}
Table 1 reports descriptive statistics including mean ($M$) and standard deviations ($SD$) for the study variables. Results of $F$ statistics in Figure 3 (a) show that participants in the High-MAST group rated Facewise significantly higher across all nine MAST criteria. In regard to trust, the difference between High-MAST and Low-MAST was significant for the \citeR{jian2000foundations} score. In addition, the High-MAST group found Facewise less risky than the Low-MAST group. Moreover, the High-MAST group rated Facewise more beneficial than those in the Low-MAST group. No significant difference in credibility ratings was found between the two conditions, possibly due to the similar presence of errors by the DSS in both levels. While the High-MAST group spent significantly more time on the task than the Low-MAST group, performance was not significantly different between the two. The High-MAST group made decisions with slightly but not significantly higher accuracy than Low-MAST. No significant differences in engagement and usability were found between High-MAST group and Low-MAST group, supporting our goal of designing both versions to be relatively equal in terms of levels of engagement and perceived usability. 

Regression analysis shows that MAST ratings are positively associated with trust ratings; people who rated trust highly also tended to rate MAST highly. Increasing the MAST score by 1 would increase one of the trust scores \cite{jian2000foundations} by 0.1 ($F(1,144)=64.94$, $p<.001$, $\beta=0.1$, $R2=0.31$) and the other trust score \cite{chancey2017trust} by 0.12 ($F(1,144)=87.83$, $p<.001$, $\beta=0.12$, $R2=0.37$). A positive relationship between MAST and credibility scores was also found; increasing the MAST score by 1 would increase credibility by 0.11 ($F(1,144)=62.96$, $p<.001$, $\beta=0.11$, $R2=0.30$). Figure 4 shows the regression plots. Furthermore, this study found that there was a negative correlation between MAST and risk score; increasing the risk score by 1 would decrease the MAST score by 2.2 ($F(1,144)=24.32$, $p<.001$, $\beta=-2.2$, $R2=0.14$). Finally, there was a positive relationship between the MAST and benefit score; increasing the benefit score by 1 would increase the MAST rating by 3.7 ($F(1,144)=71.89$, $p<.001$, $\beta=3.7$, $R2=0.33$).   

\begin{table}[H]
  \begin{center}
  \scriptsize
    \begin{tabular}{p{0.24\linewidth}|p{0.1\linewidth}|p{0.1\linewidth}|p{0.09\linewidth}|p{0.1\linewidth}|p{0.1\linewidth}|p{0.08\linewidth}}
      \toprule 
      \multirow{2}{*}{} & \multicolumn{3}{c|}{\textbf{Facewise}}  & \multicolumn{3}{c}{\textbf{READIT}} \\ 
       & \textbf{High} & \textbf{Low} & \textbf{$p$} & \textbf{High} & \textbf{Low} & \textbf{$p$} \\\midrule
       \textbf{MAST-total} & 26.5 (5.73) & 21.0 (5.77) & $<0.001*$ & 27.8 (5.38) & 19.9 (5.14) & 0.002* \\
       1. Sourcing & 2.85 (0.84) & 2.48 (0.90) & 0.011* & 3.27 (0.47) & 2.50 (1.17) & 0.05* \\
       2. Uncertainty & 2.85 (0.83) & 2.32 (0.97) & $<0.001*$ & 2.91 (0.70) & 2.42 (0.79) & 0.129 \\
       3. Distinguishing & 3.19 (0.78) & 2.15 (0.88) & $<0.001*$ & 3.27 (0.65) & 2.25 (0.87) & 0.004* \\
       4. Alternatives & 2.79 (0.82) & 2.33 (0.91) & $<0.001*$ & 2.91 (0.94) & 1.25 (0.45) & 0.001* \\
       5. Relevance & 3.00 (0.69) & 2.34 (0.89) & $<0.001*$ & 3.18 (0.75) & 3.17 (0.72) & 0.961 \\
       6. Logic & 2.97 (0.87) & 2.07 (0.96) & $<0.001*$ & 3.18 (0.87) & 2.25 (0.97) & 0.024* \\
       7. Change & 2.88 (0.83) & 2.51 (0.82) & 0.008* & 2.82 (0.75) & 1.75 (0.97) & 0.007* \\
       8. Accuracy & 2.82 (0.87) & 2.42 (0.82) & 0.005* & 3.18 (0.75) & 1.92 (0.67) & $<0.001*$ \\
       9. Visualization & 3.14 (0.75) & 2.42 (0.86) & $<0.001*$ & 3.09 (0.94) & 2.42 (0.90) & 0.095 \\
       \midrule
       \textbf{Trust (Jian)} & 4.62 (1.12) & 4.18 (1.15) & 0.023* & 5.11 (0.95) & 4.42 (1.08) & 0.119 \\
       \midrule
       \textbf{Trust (Chancey)} & 4.26 (1.28) & 4.00 (1.16) & 0.188 & 4.57 (1.25) & 3.73 (1.35) & 0.135 \\
       Chancey (Performance) & 4.24 (1.42) & 3.95 (1.28) & 0.188 & 4.85 (1.31) & 3.93 (1.47) & 0.126 \\
       Chancey (Process) & 4.68 (1.39) & 4.52 (1.36) & 0.472 & 4.76 (1.56) & 4.48 (1.53) & 0.669 \\
       Chancey (Purpose) & 3.87 (1.28) & 3.53 (1.16) & 0.093 & 4.09 (1.15) & 2.77 (1.20) & 0.013* \\
      \midrule
       \textbf{Risk} & 2.67 (1.11) & 3.10 (1.09) & 0.021* & 2.55 (0.93) & 3.33 (0.89) & 0.05* \\
       \midrule
       \textbf{Benefit} & 3.37 (0.99) & 3.01 (0.98) & 0.031* & 3.45 (0.93) & 3.00 (1.13) & 0.303 \\
       \midrule
       \textbf{Credibility} & 4.31 (1.24) & 4.23 (1.29) & 0.712 & 5.39 (0.96) & 4.44 (1.03) & 0.033* \\
       \midrule
       \makecell[l]{\textbf{Average response time} \\(seconds)} & 13.3 (4.87) & 11.3 (4.46) & 0.010* & 274 (90.7) & 214 (93.6) & 0.137\\
       \midrule
       \makecell[l]{\textbf{Performance} \\(in Platforms)} & 0.77 (0.08) & 0.75 (0.07) & \makecell[l]{0.097 \\ (accuracy)} & 2.82 (1.94) & 3.33 (1.67) & \makecell[l]{0.505 \\(report)} \\
       \midrule
       Engagement & 3.96 (1.07) & 3.99 (1.10) & 0.874 & 4.37 (0.78) & 4.52 (1.28) & 0.736 \\
       \midrule
       Usability & 3.58 (0.70) & 3.65 (0.60) & 0.494 & 3.49 (0.94) & 3.90 (0.67) & 0.248 \\
      \bottomrule 
    \end{tabular}
    \caption{Means, Standard Deviation (in parentheses), and $p$-values of study variables for High-MAST and Low-MAST groups across Facewise and READIT platforms. Asterisk(*) emphasizes the significant differences.}
    \label{tab:tableE1}
  \end{center}
\end{table}

To further validate the association between MAST and other study variables, we needed to run multiple regression analysis. However, because trust \cite{jian2000foundations,chancey2017trust}, risk, benefit, and credibility were highly correlated, it is inappropriate to run multiple regression analyses. Therefore, we applied PCA to reduce the dimensionality within our dataset. The result of PCA shows that the first two principal components explain 84.06\% of variation within the dataset. The first principal component can be perceived as an overall average of trust, risk, benefit, and credibility, while the second principal component is mainly related to negative perceptions about risk. These two principal components were used as new variables for a linear regression analysis with MAST performed for each level, High- and Low-MAST. We found that MAST-total (aggregating all MAST criteria) was highly associated with the first principal components ($F(1,144)=100.92$, $p<.001$, $\beta=0.19$, $R2=0.41$). Figure 5 provides additional details about PCA and regression results. 

\begin{figure*}[t!]
    \centering
    \begin{subfigure}[t]{0.35\textwidth}
         \centering
         \includegraphics[width=\textwidth]{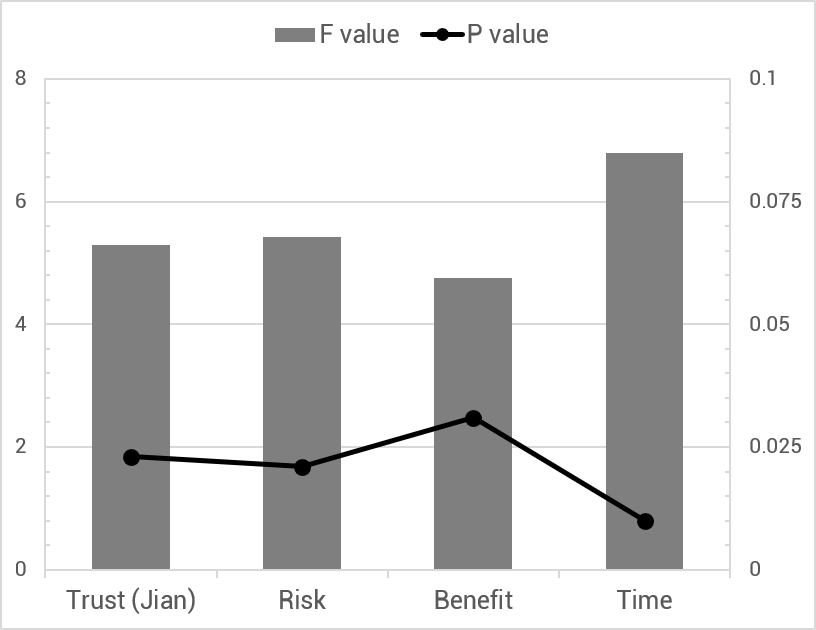}
         \caption{Facewise}
         \label{fig:8a}
     \end{subfigure}
        \begin{subfigure}[t]{0.33\textwidth}
         \centering
         \includegraphics[width=\textwidth]{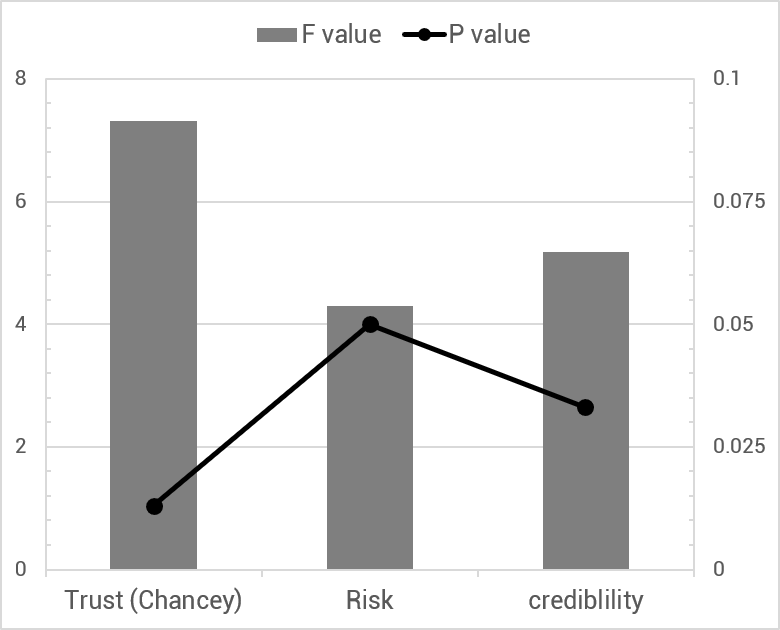}
         \caption{READIT}
         \label{fig:8b}
     \end{subfigure}
     
    \caption{The $F$-test results and corresponding $p$-values for significant variables in (a) Facewise and (b) READIT are displayed as grey bars and black dots, respectively.}
    \label{fig:fig8}
\end{figure*}

\begin{figure*}[t!]
    \centering
    \includegraphics[scale=0.4]{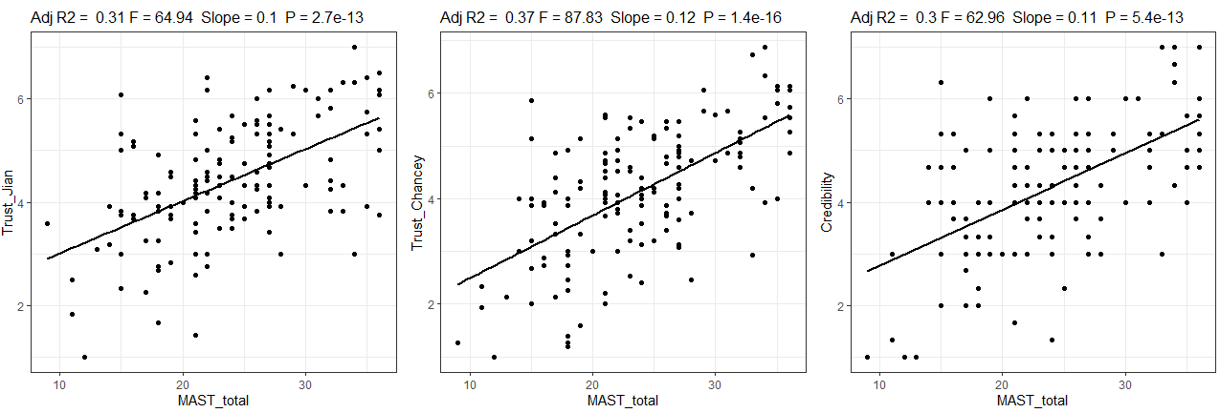}
    \caption{Least Squares Regression plots for Facewise.}
    \label{fig:fig3}
\end{figure*}

\begin{figure}[t!]
    \centering
    \begin{subfigure}[b]{0.8\textwidth}
         \centering
         \includegraphics[width=\textwidth]{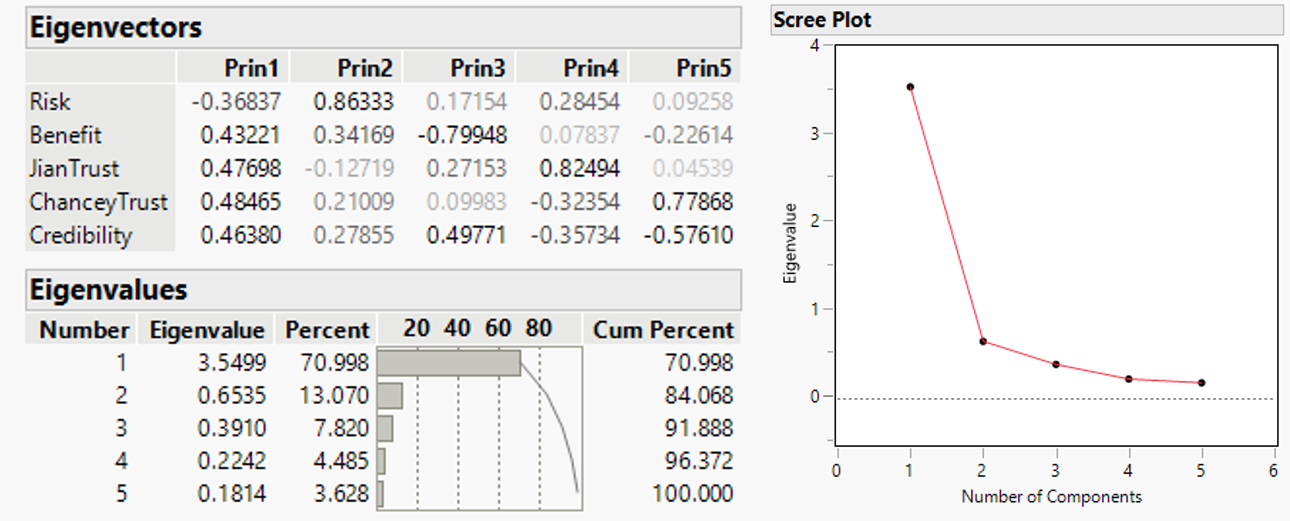}
         \caption{PCA}
     \end{subfigure}
     \hfill
     \begin{subfigure}[b]{0.8\textwidth}
         \centering
         \includegraphics[width=\textwidth]{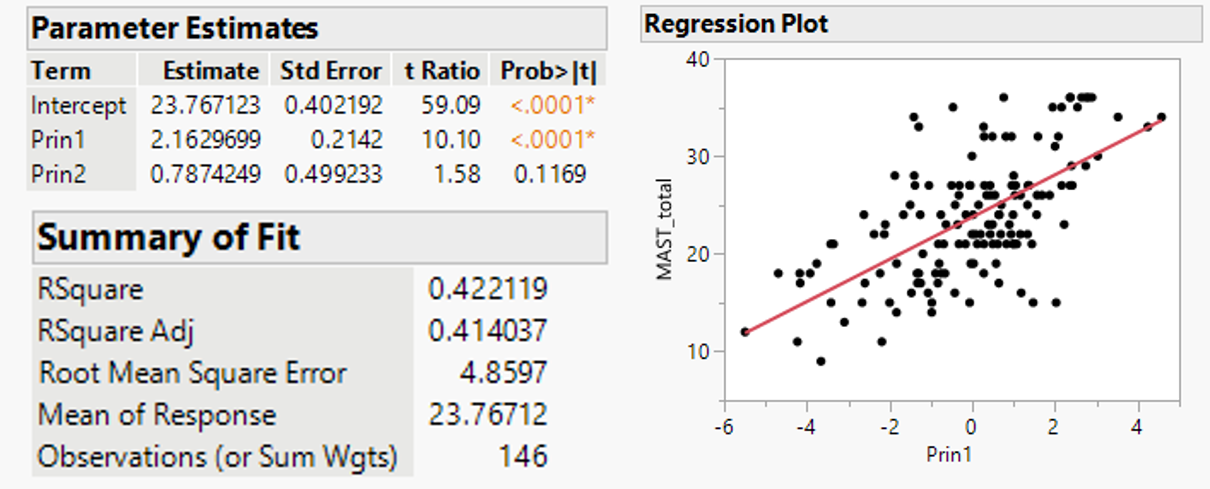}
         \caption{Linear Regression}
     \end{subfigure}
         \caption{PCA (top) and Linear regression (down) results for Facewise.}
    \label{fig:fig4}
\end{figure}
\section{Experiment 2: READIT}
\label{Experiment 2: READIT}

READIT participants were told they were intelligence analysts in a fictional major city in the United States who are tasked with monitoring the news for any ongoing threats to public safety. READIT participants were given a specific assignment to use READIT to quickly locate and search through relevant news articles and uncover a terrorist activity that had gone unnoticed for the previous five months. Figure 6 illustrates the similarities and differences between the High-MAST and Low-MAST levels of READIT. Four red dotted lines highlight the differences including the availability of “documents” and “about” tabs (Appendix D), the “topic clusters” bubble graph, the sorting option by cluster relationship strength, and the complete news pieces. For more details regarding the AI-DSS features and how they map to each of the MAST criteria, please refer to Appendix B. 

\begin{figure}[t!]
    \centering
    \begin{subfigure}{0.9\textwidth}
        \centering
        \includegraphics[width=\textwidth]{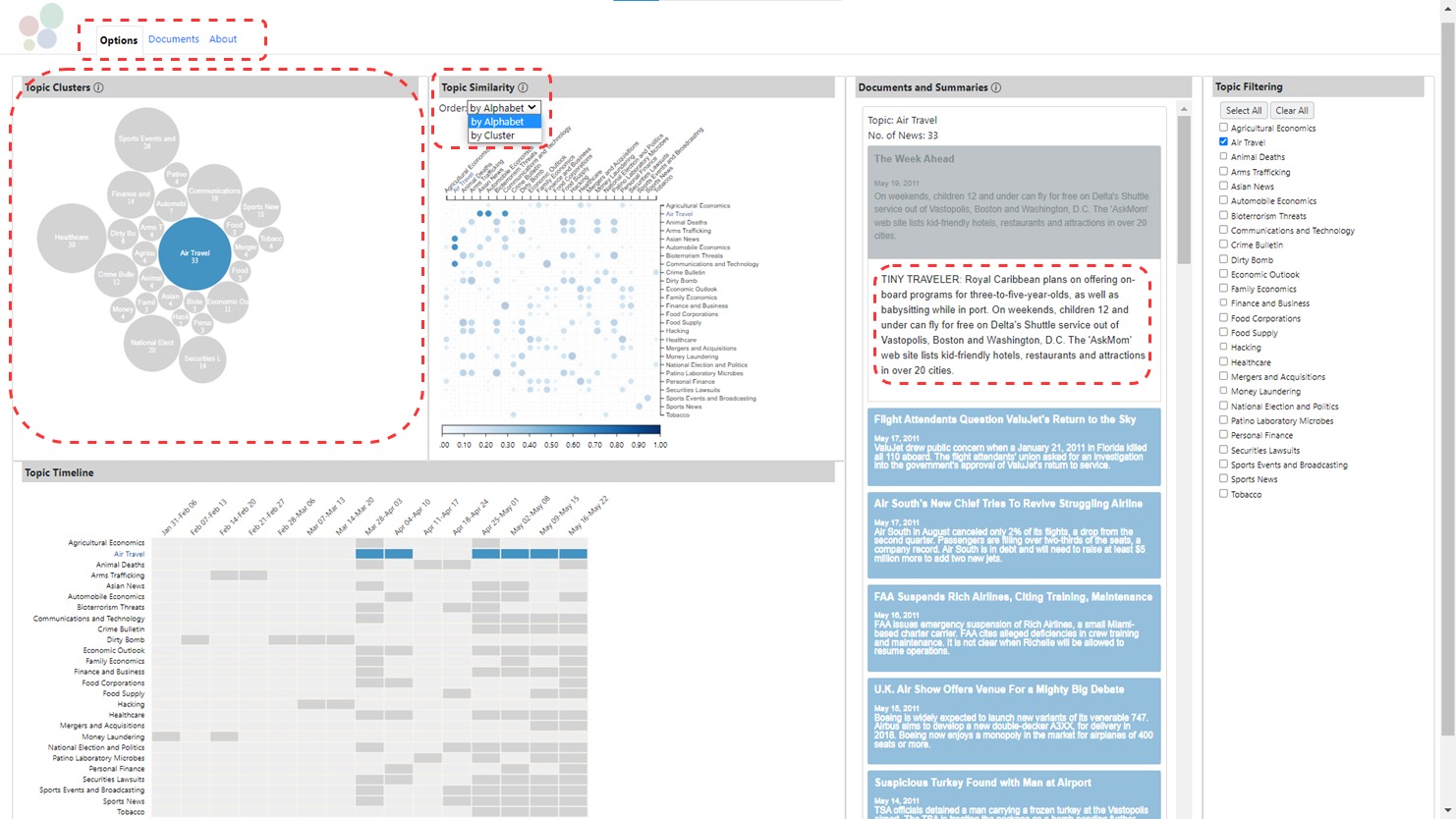}
        \caption{High-MAST READIT}
        \label{subfig:newrh}
    \end{subfigure}
    \begin{subfigure}{0.9\textwidth}
        \centering
        \includegraphics[width=\textwidth]{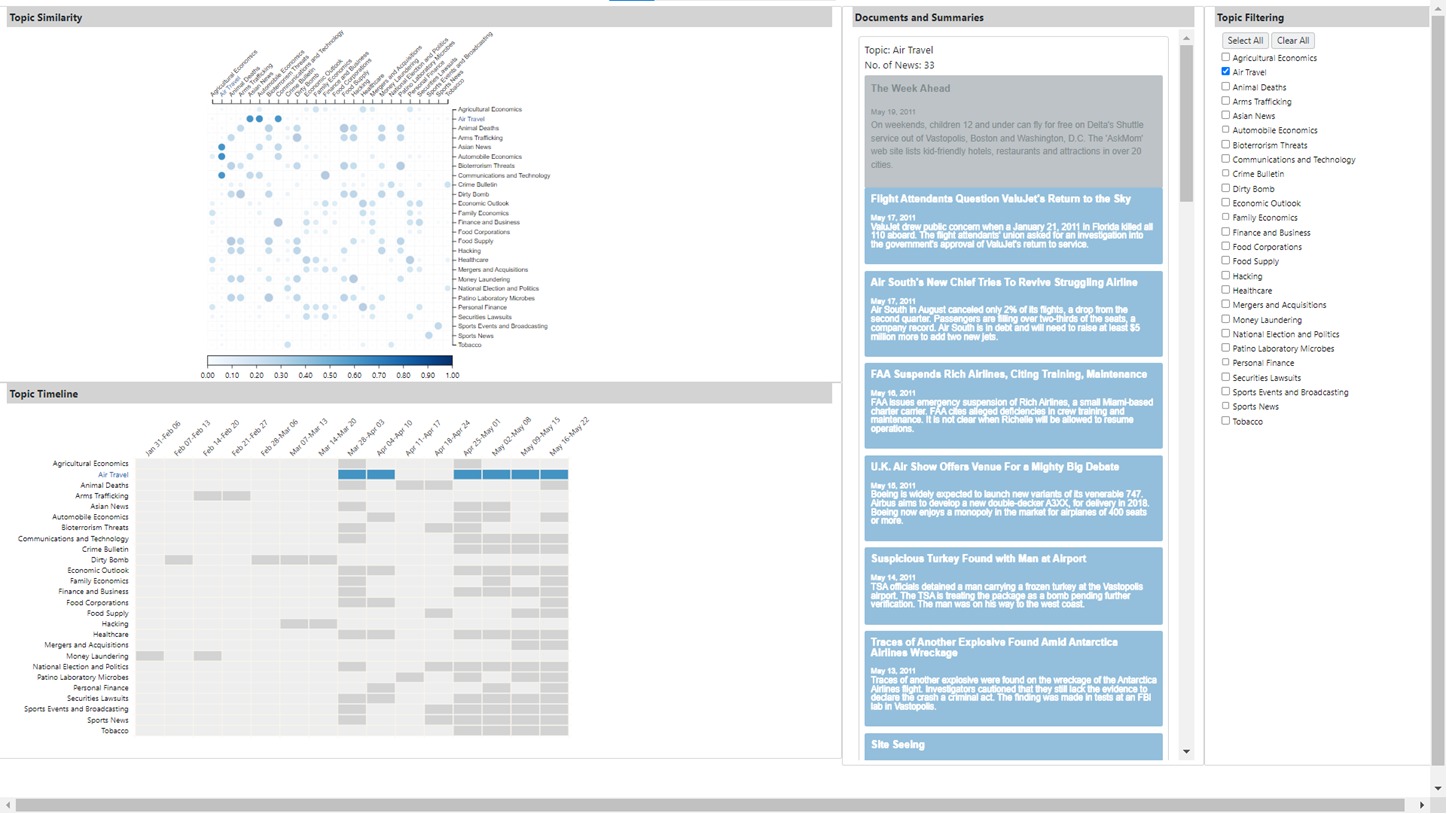}
        \caption{Low-MAST READIT}
        \label{subfig:newrl}
    \end{subfigure} 
    \caption{High-MAST READIT (top) and Low-MAST READIT (down). Compared with the Low-MAST READIT, High-MAST READIT has more interactive features (Topic Clusters, Topic Similarity, original documents, clickable timelines, etc.) to demonstrate the MAST criterion.}
    \label{fig:fig5}
\end{figure}

\subsection{Participants}
A total of 25 Intelligence Analysts (IAs) from the U.S. Department of Homeland Security (DHS) were recruited to complete our study, administered through Microsoft Teams or Zoom, over a period of 19 days. Two participants were unable to complete the study due to unexpected scheduling restrictions. The resulting High-MAST and Low-MAST versions of READIT were tested with a sample of 11 and 12 IAs, respectively. On average, participants spent 75 minutes to complete the study, including onboarding and responses to questionnaire items. We were not permitted to compensate participants monetarily because they were federal employees. However because participants were self-selected volunteers who responded to our recruitment script and were willing to spend time completing our study, we assumed they were sufficiently motivated to complete this study to the best of their ability. Table 6 in Appendix F reports the participant demographics per condition.

\subsection{Results and Discussion}
Table 1 reports descriptive statistics including mean and standard deviations for the study variables. Results show that the High-MAST group rated READIT significantly higher on the MAST criteria than the Low-MAST group for six out of nine MAST criteria (i.e., except for uncertainty, relevance, and visualization). Trust ratings were also generally higher for those in the High-MAST group; however, the difference is only significant for the purpose dimension of the \cite{chancey2017trust} trust score. Moreover, the High-MAST group compared to the Low-MAST group found READIT significantly less risky to use, and more credible. No significant differences in performance were found between High-MAST and Low-MAST groups in terms of average response time, or on their 250-word report, although descriptively the Low-MAST group spent less time completing the task and had higher performance scores than the High-MAST group. No significant differences were found in engagement and usability ratings between High-MAST and Low-MAST groups, supporting our original intention to keep the different READIT versions roughly equivalent in terms of levels of engagement and usability. Figure 3 (b) shows the \textit{F} values of significant variables in READIT.

Regression analysis showed that MAST ratings are positively associated with trust ratings. There is a positive relationship between MAST and both the trust scores \cite{jian2000foundations,chancey2017trust}; increasing MAST by 1 increases one of the trust scores \cite{jian2000foundations} by 0.13 ($F(1,21)=32$, $p<.001$, $\beta=0.13$, $R2=0.58$) and increases the other trust score \cite{chancey2017trust} by 0.16 ($F(1,21)=35.29$, $p<.001$, $\beta=0.16$, $R2=0.61$). A positive relationship was also found between MAST and credibility scores; increasing the MAST score by 1 would increase credibility ratings by 0.14 ($F(1,21)=52.46$, $p<.001$, $\beta=0.14$, $R2=0.70$). Figure 7 shows the regression plots. This study also found that there was a negative relationship between MAST and risk score; increasing the MAST score by 1 would decrease perceived risk by 4.9 ($F(1,21)=24.89$, $p<.001$, $\beta=-4.9$, $R2=0.52$). In addition, there was a positive relationship between MAST and perceived benefit; increasing the benefit score by 1 would increase MAST by 4.2 ($F(1,21)=17.19$, $p<.001$, $\beta=4.2$, $R2=0.42$). 

\begin{figure*}[t!]
    \centering
    \includegraphics[scale=0.4]{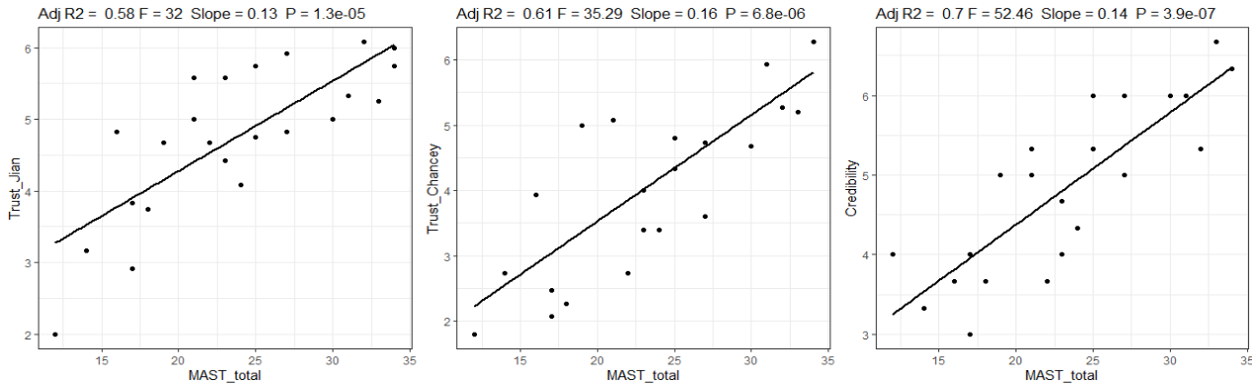}
    \caption{Least Squares Regression plots for READIT.}
    \label{fig:fig6}
\end{figure*}

Because trust \cite{jian2000foundations,chancey2017trust}, risk, benefit, and credibility were highly correlated for the READIT platform, we did not run multiple regression and instead used PCA. The PCA results show that the first two principal components can explain 87.66\% of variation within the dataset. The first principal component can be interpreted as an overall average of trust, risk, benefit, and credibility. However, the second principal component was primarily related to negative perceptions about risk. For all observations, two PC scores were calculated using each principal component and these scores served as the regressors for further analysis. We found that averaging across all MAST criteria to produce a MAST-total score can significantly predict the first principal components ($F(1,144)=60.07$, $p<.001$, $\beta=0.26$, $R2=0.74$). Figure 8 provides more details about the PCA and regression results. 

\begin{figure*}[ht]
    \centering
    \begin{subfigure}[b]{0.8\textwidth}
         \centering
         \includegraphics[width=\textwidth]{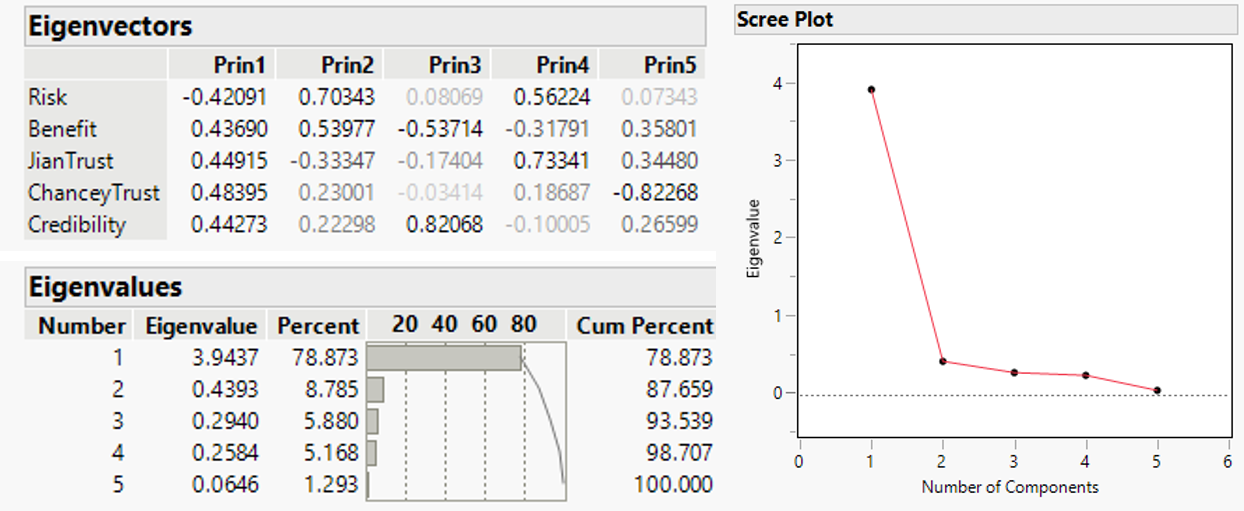}
         \caption{PCA}
     \end{subfigure}
     \hfill
     \begin{subfigure}[b]{0.8\textwidth}
         \centering
         \includegraphics[width=\textwidth]{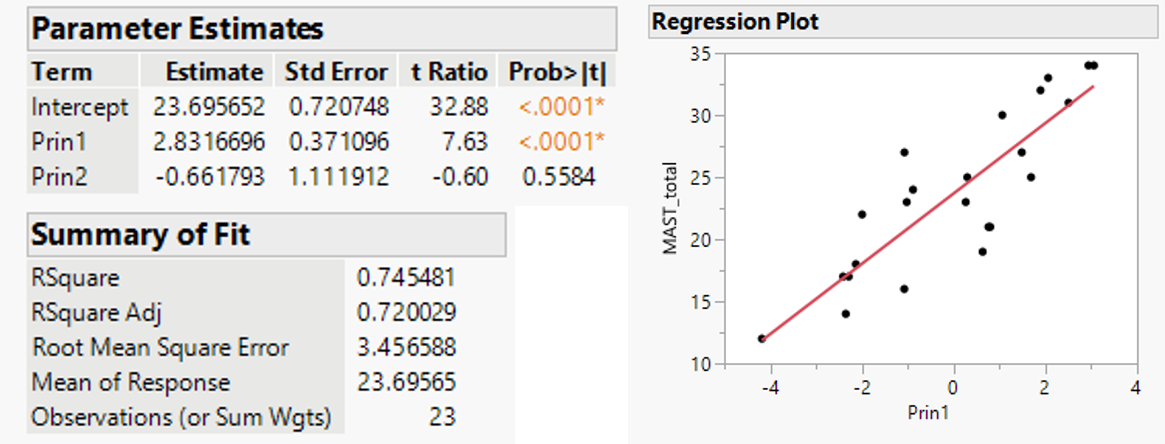}
         \caption{Linear Regression}
     \end{subfigure}
     
    \caption{PCA (top) and Linear regression (down) results for READIT.}
    \label{fig:fig7}
\end{figure*}

\section{General Discussion}
\label{General Discussion}

For both Facewise and READIT, our results show that our High-MAST conditions achieved higher MAST ratings for Facewise (9/9 criteria for Facewise and 6/9 criteria for READIT). This disparity highlights the significant role that the type of system plays in operationalizing MAST criteria. Facewise involves relatively straightforward operational criteria, where factors such as accuracy, source reliability, and user interface clarity are more directly applicable and measurable. This likely accounts for the uniform high performance across all criteria for Facewise in High-MAST conditions. Conversely, READIT, a text summarization system, presents unique challenges. The inherent complexity of natural language processing and the subjective nature of text interpretation make operationalizing certain MAST criteria more difficult. Additionally, the visualization of results in text summarization does not show a distinct difference between High- and Low-MAST conditions, with three main visualizations in High-MAST and two in Low-MAST. This narrows the difficulty gap between the two versions.

Results also show that High-MAST conditions had higher trust ratings (\citeR{jian2000foundations} for Facewise and the ``purpose'' dimension from \citeR{chancey2017trust} for READIT) and lower perceived risk scores compared to the Low-MAST conditions. Benefit and credibility ratings were also higher for both High-MAST groups. However, differences in benefit and credibility ratings were significant only for Facewise and READIT, respectively. In both platforms, usability and engagement were equally rated by Low-MAST and High-MAST groups. No clear trends for performance were found for either DSS. Regression analysis results suggest that people who rated MAST criteria highly tended to rate trust \cite{jian2000foundations,chancey2017trust}, credibility, and benefit high, and risk low. In addition, the applied principal component analysis found that MAST-total ratings can be significantly explained by trust, credibility, risk, and benefit responses.

Lastly, results from the \citeR{jian2000foundations} questionnaire show a statistically significant difference for Facewise but not for the \citeR{chancey2017trust} questionnaire. We speculate that this might be due to the nature of the \citeR{jian2000foundations} questionnaire items being valenced both negatively and positively whereas for the \citeR{chancey2017trust} questionnaire, the items are all positively valenced. Prior research has shown that item valence can affect responses in trust measures \cite{doi:10.1177/1071181319631201}, and that negatively valenced trust items may result in more variable responses compared to positively valenced trust items \cite{Schroeder_Chiou_Craig_2021}. More research is needed to investigate  why these two instruments  measuring the same construct could result in different responses \cite{Long_Sato_Millner_Loranger_Mirabelli_Xu_Yamani_2020}. 

To address the criticism that our DSSs designed to score high on the MAST tool will naturally score high when using MAST as an evaluation tool, it is crucial to understand the different types of validation methods and the need for continued research in this area. By demonstrating that AI-DSSs designed to align with the MAST ratings then indeed score respectively so in a human subjects study (rather than an assessment conducted by the researchers as evaluators), the study thus helps to validate that the MAST criteria are generally measuring what it was intended to measure. Establishing internal validity of such tools is an initial step towards a more open and collective effort to validate the tool externally, ecologically, and provides a reference point for future development of the tool involving the broader researcher and practitioner communities. 

\subsection{Limitations}
There are several limitations to this study. First, although we tried to create realistic and useful AI-DSS accounting for the MAST criteria, our resulting DSSs in either High-MAST or Low-MAST versions ultimately did not result in significant performance differences. Some possible reasons could be a lack of meaningful variation in the image database and under-optimized AI algorithms used for Facewise that were shared across both versions of the DSS. For READIT, although we addressed this during the onboarding video, it is possible that some people may have confused the size of the bubble graph topic clusters as communicating degrees of importance rather than degrees of normality in a task that essentially required searching for anomalies. From a design perspective, it was a challenge with the limited time we had to develop the READIT interface to balance the screen real estate on a single browser page that would get across the right level of detail in an easy to navigate way without leading participants too easily to the ground truth and causing a ceiling effect with performance. Future studies can develop these AI-DSS testbeds further by improving upon these study designs including AI performance, the databases used, and more systematically optimizing the level of information detail for each interface. 

Second, while MAST ratings were highly associated with trust, our results do not factor in whether trust or distrust levels were calibrated with system performance. Such an analysis may be possible for Facewise, in which system reliability can be precisely gauged using signal detection metrics. In contrast, trust and distrust calibration is difficult to precisely define for the READIT platform because it does not offer direct recommendations or answers that could be rated as easily. Future studies may want to consider these different forms of decision support \cite{chiou2023trusting}, and how those different forms can affect trust responses. Third, although the signals were very strong for READIT, we could not reach our intended sample size due to the challenge of recruiting expert IAs as volunteer participants in a remote study. Lastly, this study only used MAST as a framework for designing our AI-DSSs because of our study's primary objectives, and a comprehensive review and comparison of MAST and other similar frameworks would be beyond the scope of this project. However, other literature has reviewed similar tools for trust assessment \cite{Kohn_deVisser_Wiese_Lee_Shaw_2021,Alsaid_Li_Chiou_Lee_2023}. MAST might be used in conjunction with some of these other tools to achieve a more comprehensively designed system, and possibly better performance outcomes. 

\section{Conclusion and Future Directions}
\label{Conclusion and Future Directions}

The primary objective of this study was to establish the utility of the Multisource AI Scorecard Table (MAST) for evaluating perceived trustworthiness in AI-enabled decision support systems. This resulted in an interesting opportunity to evaluate whether the tradecraft standards behind MAST are related to the existing tools developed by the scientific community of trust researchers. The results of our study show strong associations between MAST and trust assessments across two domains of application. While MAST was initially conceptualized for intelligence type reporting tools such as READIT, we showed that these patterns of associations persist with automated face recognition systems as well. This indicates that in these types of information processing and cognitive decision-making domains, MAST may be a versatile tool in designing \cite{blasch2021multisource} and evaluating perceptions of AI system trustworthiness. 

In comparison to other guidelines and frameworks for evaluating AI-enabled systems, the benefit of using MAST is that it has been operationalized as a rating system by a practitioner community \cite{blasch2021multisource}. Thus, for its intended operational use, we had known that its basic constructs and the values behind those constructs were understandable to and accepted by the intelligence community. By testing the tool on airport security officers, we have also demonstrated its operational value in other high-stakes domains. Lastly, as an evaluative instrument, we also showed that the tool is well-aligned with the scholarly and science-supported constructs of trust and credibility.

It is important to note that High-MAST ratings do not necessarily translate to improved performance in the human-AI system, a finding supported in previous work. There are a variety of factors that contribute to this, including factors in the task environment, mental workload, and task difficulty \cite{sargent2023metaanalysis}. Additional analysis addressing these factors in the task environment along with a more in-depth exploration of the behavioral data captured during the interactive task will be needed to understand the causal mechanisms that underpin this phenomenon.


\acks{This material is based on work supported by the U. S. Department of Homeland Security under Grant Award Number 17STQAC00001-05-00. The views and conclusions contained in this document are those of the authors and should not be interpreted as representing the official policies, either expressed or implied, of the Department of Homeland Security. PS contributed the initial complete draft of the paper. YB contributed to data analysis. NK, MC, YB, SB contributed written sections and response to reviewers. NK, AP, AM, YW, JZ contributed to design and development of Facewise and READIT. PS, AP, and MC contributed other study materials. PS, AP, MC, YB, NK, and AM contributed to data collection. MM, JS, EB contributed to study conception, design, interpretation, and participant recruitment. EC contributed to all aspects of the study. All authors reviewed and approved the final version of this paper.
}

\bibliography{refs}
\bibliographystyle{theapa}

\newpage 
\appendix
\section*{Appendix A. The Nine MAST Criteria for Facewise}
\label{appendixA}

\small

\begin{longtable}{p{0.2\linewidth}|p{0.7\linewidth}}
    
    \toprule
    \textbf{MAST item} & \textbf{Questions and Feature Descriptions} \\
    \midrule
    \multirow{3}{*}{\textbf{Sourcing}} & How well can the system identify underlying sources and methodologies upon which results are based?  \\
                                      & \textbf{High-MAST:} The ``View Details'' page provides the name of image sources and demographical information about the people whose image data were used to train the AI, such as their race and gender.\\ 
                                      & \textbf{Low-MAST:} The system interface does not include the name of image sources and demographical information about the people whose image data were used to train the AI system, such as their race and gender.\\
    
      \midrule 
      \multirow{3}{*}{\textbf{Uncertainty}} & How well can the system indicate and explain the basis for the uncertainties associated with derived results? \\
                                            & \textbf{High-MAST:} For each pair of images, the system will display a certainty score from 0\%-100\% to indicate its confidence about its recommended decision. The system also gives an alert if the uncertainty is too high when you click the “Final Decision” button, depending on your decision. Details about how the system calculates the certainty score are available by clicking on the “More Details” button under every decision. The AI's confidence level is calculated in this manner: first, a metric that signifies the mathematical distance between two image pairs is calculated. Then, the difference between the mathematical distance and a pre-determined (computed during the training and validation stages) threshold is calculated. Finally, the difference is normalized by a factor and the confidence level is calculated using probability measures associated with the standard normal distribution. Thus, the AI's confidence is an indication, based on the predetermined threshold. Confidence levels closer to 100\% indicate higher confidence.\\
                                            & \textbf{Low-MAST:} For each pair of images, the system only recommends a binary decision (same or different) and does not indicate its confidence in the decision.\\

      \midrule 
      \multirow{3}{*}{\textbf{Distinguishing}}  & How well can the system clearly distinguish derived results and underlying data?\\
                                                & \textbf{High-MAST:} The system can distinguish whether a presented ID is invalid or expired, or if the ID photo may have been digitally altered. An alert message will be automatically shown in these cases by the system. Details about how the system identifies these features in the ID photo are available by hovering over the Crossmark or checkmark icon next to the ID expiration date.\\
                                                & \textbf{Low-MAST:} The system cannot distinguish whether a presented ID is invalid or expired, or if the ID photo may have been digitally altered. \\
      \midrule 
      \multirow{3}{*}{\textbf{\makecell{Analysis of \\ Alternatives}}}    & How well can the system identify and assess plausible alternative results?\\
                                                            & \textbf{High-MAST:} In the ``View Details'' page, the system provides dissimilarity and similarity probabilities as alternatives for each pair. The similarity and dissimilarity numbers are directly derived from the AI's confidence level. The higher of the two probabilities is selected to represent the AI's confidence level. The calculation of the similarity and dissimilarity probabilities assumes that the threshold is distributed as standard normal, and that the scaled differences are realizations of a noise-generating process. Both probabilities are calculated using the scaled difference between the distance metric and the threshold.\\
                                                            & \textbf{Low-MAST:} For each pair of images, the system only gives a decision and does not indicate its confidence in the current decision based on the training and validation stages, nor on probability measures of alternatives associated with the standard normal distribution.\\
      \midrule 
      \multirow{3}{*}{\textbf{\makecell{Customer \\Relevance}}}  & How well can the system provide information and insight to users?\\
                                                    & \textbf{High-MAST:} Besides providing the binary decision of same or different, the confidence level, and ID validation on the main page, the system provides additional details through a "More Details" button. This includes information and explanations about similarity, dissimilarity, confidence level, and sources of training for AI. To present the information more efficiently, the system will minimize explanations that have already been shown. Conditional alerts when the system’s certainty level is low and alerts about individuals who may need additional screening per the protocol are also included as part of the system with the information displayed as detected.\\
                                                    & \textbf{Low-MAST:} Besides providing the binary decision of the same or different and ID expiration date on the main page, the system does not provide any additional details or any conditional alerts.\\
      \midrule 

      \multirow{3}{*}{\textbf{Logic}}   & How well can the system help the user understand how it derived its results?\\
                                        & \textbf{High-MAST:} The system bases its final decision by choosing the larger of similarity and dissimilarity probabilities. “More Details” button also provides an explanation and interpretation of how a prediction or classification is made. Conditional alerts when the system’s certainty level is low, and alerts about individuals who may need additional screening per the protocol are also included. To detect the authenticity of an ID photo, a second model was trained, tested, and validated on proprietary datasets of anomalous and non-anomalous travel documents, digitally altered and original images. A separate model further performs character recognition to analyze expiration dates on travel documents.\\
                                        & \textbf{Low-MAST:} The system does not give any information on how its recommendation is determined. It also does not provide any conditional alerts or any information about the authenticity or validity of the ID photo image.\\    
      \midrule 
      \multirow{3}{*}{\textbf{Change}}  & How well can the system help the user understand how derived results on a topic are consistent with or represent a change from previous analysis of the same or similar topic?\\
                                        & \textbf{High-MAST:} As you interact with the system, by clicking “more details” you will see a report about your agreement with the system, which indicates how often the system has been uncertain about your final decisions.\\
                                        & \textbf{Low-MAST:} As you interact with the system, the system does not indicate how often it has been uncertain about your final decisions.\\
      \midrule 
      \multirow{3}{*}{\textbf{Accuracy}}    & How well can the system make the most accurate judgments and assessments possible, based on the information available and known information gaps?\\
                                            & \textbf{High-MAST:} For each pair of images, the system will display a certainty score from 0\%-100\% to indicate its confidence about its recommended decision. System's performance according to the training data and more details about how the system calculates the certainty score are available by clicking on the “More Details” button under every decision.\\
                                            & \textbf{Low-MAST:} For each pair of images, the system only gives a binary decision and does not indicate its confidence in the decision, the system's performance according to the training data, or more details about how the system made the decision.\\
      \midrule 

      \multirow{3}{*}{\textbf{Visualization}}   & How well can the system incorporate visual information if it will clarify an analytic message and complement or enhance the presentation of data and analysis? Is visual information clear and pertinent to the product’s subject?\\
                                                & \textbf{High-MAST:} The system automatically shows you an enlarged version of a traveler's ID photo and their photo taken at the security checkpoint. These images will be shown side by side. Distinguishing features that played a big role in determining the recommended decision will also be highlighted by clicking the ``View Details'' button.\\
                                                & \textbf{Low-MAST:} The system only shows you an enlarged version of a traveler's ID photo and their photo taken at the security checkpoint without any additional visualized explanation about the recommended decision. \\
    \bottomrule
    \caption{MAST \cite{blasch2021multisource} and Facewise feature descriptions for High-MAST and Low-MAST.}
    \label{tab:tableA1} \\
\end{longtable}

\newpage 
\section*{Appendix B: The Nine MAST Criteria for READIT}
\label{appendixB}

\small

\begin{longtable}{p{0.2\linewidth}|p{0.7\linewidth}}

    \toprule
    \textbf{MAST item} & \textbf{Questions and Feature Descriptions} \\
    \midrule
    
    
    \multirow{3}{*}{\textbf{Sourcing}} & How well can the system identify underlying sources and methodologies upon which results are based?  \\
                                      & \textbf{High-MAST:} In the documents page, you can see descriptive information about the documents (data) used to gather the clusters including basic information and detailed descriptions of the sources. The datasheet for READIT includes information on the clustering model, models for summarization, training data, possible biases, pre-processing of data, and quality of the data used in training to derive results. In the main dashboard view, you can view the data used to derive the cluster either by hovering or clicking on it including the cluster title, number of documents, top terms, and representative documents. The representative documents can be viewed as a summary (derived result) or raw version.\\ 
                                      & \textbf{Low-MAST:} For any given cluster in the main dashboard view, you can view more details about it by clicking on it. The title of the cluster, number of documents, and summaries of the documents will be displayed in the documents and summaries pane. Only the derived results are shown, not the underlying sources and data used to derive the clusters or summaries.\\
    
      \midrule 
      \multirow{3}{*}{\textbf{Uncertainty}} & How well can the system indicate and explain the basis for the uncertainties associated with derived results? \\
                                            & \textbf{High-MAST:} READIT indicates levels of uncertainty with derived results in two ways, as described in the datasheet. First, READIT includes keywords per cluster to show how documents in clusters are related to each other. Keywords are displayed with a term frequency–inverse document frequency (tf-idf) score which measures the certainty the word fits with the cluster. Second, READIT includes similarity scores to assess the similarity between clusters. This score is calculated using cosine similarity to show the certainty that clusters are related to each other.\\
                                            & \textbf{Low-MAST:} In the topic similarity visualization, the relationship between two topics is colored from white to dark blue with dark blue indicating a higher certainty the two topics are related. These relationships are not labeled with numbers, neither is it explained how this similarity is calculated.\\

      \midrule 
      \multirow{3}{*}{\textbf{Distinguishing}}  & How well can the system clearly distinguish derived results and underlying data?\\
                                                & \textbf{High-MAST:} For any given cluster you can view more details about the data used to derive the cluster either by hovering or clicking on it. The datasheet for READIT includes information on the clustering model, models for summarization, training data, underlying assumptions for choice of training data, quality of the data used in training to derive results, possible biases, pre-processing of data, recommended uses and users, and restrictions on use. The datasheet was created with domain expert input.\\
                                                & \textbf{Low-MAST:} When opening or clicking on clusters, you can view more details about that cluster. The title and summary of representative documents will appear. The raw data used to derive the title and summaries is not displayed. There is no datasheet with information on how these titles or summaries are calculated. \\
      \midrule 
      \multirow{3}{*}{\textbf{\makecell{Analysis of \\Alternatives}}}    & How well can the system identify and assess plausible alternative results?\\
                                                            & \textbf{High-MAST:} In the topic similarity visualization, users initially view the visualization where the topics are ordered alphabetically. By factoring in the similarity score and uncertainties, READIT can reorder the view in this visualization such that highly related topics will appear together to present an alternative view.\\
                                                            & \textbf{Low-MAST:} READIT is not able to show alternative results when uncertainties in the data warrant them. There is no way to reorder visualizations based on any criteria.\\
      \midrule 
      \multirow{3}{*}{\textbf{\makecell{Customer \\Relevance}}}  & How well can the system provide information and insight to users?\\
                                                    & \textbf{High-MAST:} READIT synthesizes large corpora of documents and produces clusters of similar documents. The topic similarity visualization shows which clusters are most highly related to each other. Users can examine the clusters and their relationships in the topic similarity view for trends for follow-up work. READIT is also able to suggest locations to filter by if the documents contain multiple locations. Users can also filter all visualizations by topic. There is a topic filtering pane where users can check all, or some topics and the corresponding selected topics will be highlighted in the visualizations.\\
                                                    & \textbf{Low-MAST:} READIT synthesizes large corpora of documents and produces clusters of similar documents. The topic similarity visualization shows which clusters are most highly related to each other. Users can examine the clusters and their relationships in the topic similarity view for trends for follow-up work.\\
      \midrule 

      \multirow{3}{*}{\textbf{Logic}}   & How well can the system help the user understand how it derived its results?\\
                                        & \textbf{High-MAST:} For any given cluster you can view more details about the data used to derive the cluster either by hovering or clicking on it. The datasheet includes information on pre-processing of data. READIT includes an option to filter results by location, if location information is detected in the document. To give location options, READIT must consider the location information in the context of the document, and other assumptions about the embedding of the location in the document.\\
                                        & \textbf{Low-MAST:} When clicking on clusters in the main view, you can view the title and representative documents in summary form. The titles and summaries are understandable to users. Information on how clusters, titles, and summaries are formed is not included. There is also no information on the pre-processing of data.\\    
      \midrule 
      \multirow{3}{*}{\textbf{Change}}  & How well can the system help the user understand how derived results on a topic are consistent with or represent a change from previous analysis of the same or similar topic?\\
                                        & \textbf{High-MAST:} In the documents page, READIT includes information on similar searches from other agencies. Similar searches may be based on the average length of the document, number of documents, or number of clusters generated.\\
                                        & \textbf{Low-MAST:} READIT does not have a way to note changes from previous analyses or similar analyses. READIT also cannot compare current results with those of other agencies which had similar results.\\
      \midrule 
      \multirow{3}{*}{\textbf{Accuracy}}    & How well can the system make the most accurate judgments and assessments possible, based on the information available and known information gaps?\\
                                            & \textbf{High-MAST:} The READIT datasheet includes information on system verification and validation methodology, and results from the training data where the system achieved sufficiently high accuracy. To assess the accuracy of READIT, users can view the full documents used in each cluster and compare them against the top terms to independently determine whether the documents match the top terms. Likewise, users can view a summary of the document and compare it against the full version of the document in the documents and summaries view to see if the summary is accurate.\\
                                            & \textbf{Low-MAST:} READIT does not include information on system verification, validation methodology, or information on the training of the system where it achieved sufficient accuracy. Since underlying sourcing information and raw data are not included in the system, it is difficult to assess whether the topics and summaries are accurate.\\
      \midrule 

      \multirow{3}{*}{\textbf{Visualization}}   & How well can the system incorporate visual information if it will clarify an analytic message and complement or enhance the presentation of data and analysis? Is visual information clear and pertinent to the product’s subject?\\
                                                & \textbf{High-MAST:} READIT uses three main visualizations to enhance users’ understanding of the clusters. First, in the topic overview visualization, clusters are displayed as bubbles where the size of the bubbles can indicate anomalies. Next, READIT also creates and displays a topic similarity visualization to help understand the connections between clusters. Lastly, there is a timeline view in READIT to display clusters on a timeline (if documents contain date information). All visualizations are simple and labeled properly. Users can view more details about the visualizations by clicking on them or hovering over them or filtering all visualizations by cluster using the filtering option.\\
                                                & \textbf{Low-MAST:} READIT uses two visualizations. The similarity matrix shows the similarity scores between topics. Darker colors indicate more similarity but score values are not shown. The timeline shows the clusters on the timeline. Visualizations contain no interactivity and users are not able to click or hover on items to view more details about the visualizations. \\
    \bottomrule 
    \caption{MAST \cite{blasch2021multisource} and READIT feature descriptions for High-MAST and Low-MAST.}
    \label{tab:tableB1} \\
\end{longtable}

\newpage 
\section*{Appendix C: Study Questionnaires}
\label{appendixC}

\scriptsize

\begin{longtable}{p{0.12\linewidth}|p{0.12\linewidth}|p{0.2\linewidth}|p{0.1\linewidth}|p{0.08\linewidth}|p{0.1\linewidth}|p{0.1\linewidth}}
    \toprule
    \textbf{Variables} & \textbf{Reference} & \textbf{Example item(s)} & \textbf{Number of items/ Reverse items} & \textbf{Scale} & \textbf{Facewise Cronbach’s Alpha} & \textbf{READIT Cronbach’s Alpha}\\
    \midrule
    \textbf{MAST-total} & (Blasch et al., 2020) & Sourcing, uncertainty, distinguishing, analysis of alternatives, customer relevance, logical argumentation, consistency, accuracy, and visualization & 9/0 & 9 - 36 & .91 & .91\\
    
      \midrule 
      \textbf{Risk} & (Weber et al., 2002) & Please indicate how risky you perceive it is to use this system for completing your task well. & 1/0 & 1 - 5 & - & -\\
      \midrule 
     \textbf{Benefit} & (Weber et al., 2002) & Please indicate how beneficial you perceive it is to use this system for completing your task well. & 1/0 & 1 - 5 & - & -\\
      \midrule 
      \textbf{\makecell[l]{Trust \\(Jian)}} & (Jian et al., 2000) & “I can trust the system.”; “The system looks deceptive.” & 12/5 & 1 - 7 & 0.90 & 0.92\\
      \midrule 
      \textbf{Trust (Chancey)} & (Chancey et al., 2017) & “I understand how the system will help me perform well. “; “The information the system provides reliably helps me perform well. & 15/0 & 1 - 7 & 0.96 & 0.96\\
      \midrule 

      \textbf{Credibility} & (Appelman \& Sundar, 2016) & “How accurate do the results of the system appear to be?”; “How believable do the results of the system appear to be?” & 3/0 & 1 - 7 & 0.92 & 0.92\\
      \midrule 
      \textbf{Engagement} & (Schaufeli et al., 2002) & “I was immersed in this research task.”; “To me, this research task was challenging.” & 17/0 & 1 - 7 & 0.91 & 0.93\\
      \midrule 
      \textbf{Usability (SUS)} & (Brooke, 2020) & “I felt very confident using the system.”; “I thought the system was easy to use.” & 10/5 & 1 - 5 & 0.80 & 0.88\\
      \midrule 
      \textbf{\makecell[l]{Task \\performance}} & - & Average response time and Accuracy for Facewise and final report gradings for READIT & 2/0 & 0 & - & -\\
      \bottomrule
  \caption{Dependent and Control Variables.}
    \label{tab:tableC1} \\
\end{longtable}

\newpage 
\section*{Appendix D: READIT Documents and About Tabs for the High-MAST Version}
\label{appendixD}

\begin{figure*}[ht]
    \centering
    \begin{subfigure}[t]{0.49\textwidth}
         \centering
         \includegraphics[width=\textwidth]{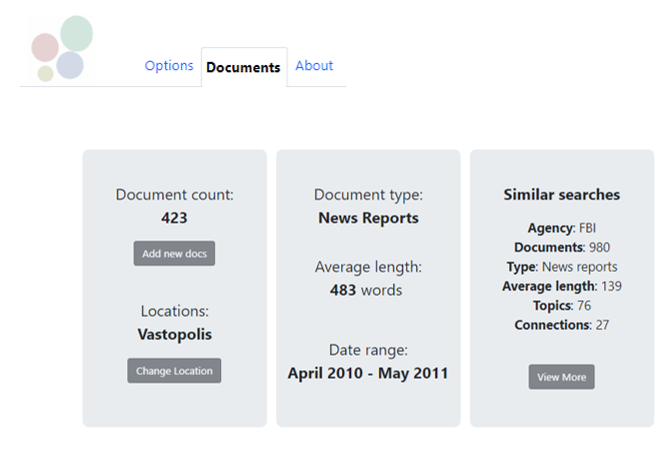}
         \caption{Documents tab}
     \end{subfigure}
     \begin{subfigure}[t]{0.43\textwidth}
         \centering
         \includegraphics[width=\textwidth]{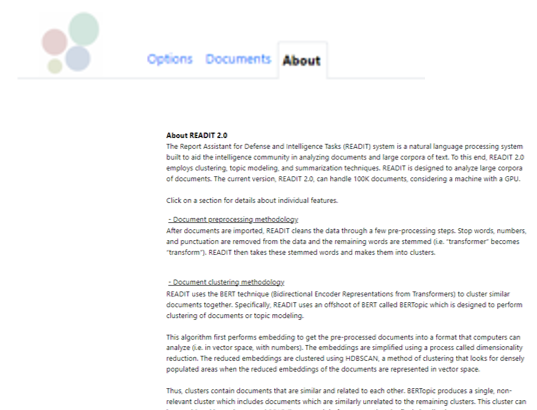}
         \caption{About tab}
     \end{subfigure}
     
    \caption{Documents and About tabs in High-MAST READIT platform.}
    \label{fig:figD}
\end{figure*}

\newpage 
\section*{Appendix E: 95\% Confidence Interval Figures}
\label{appendixE}

\begin{figure}[h]
    \centering
    \includegraphics[scale=0.5]{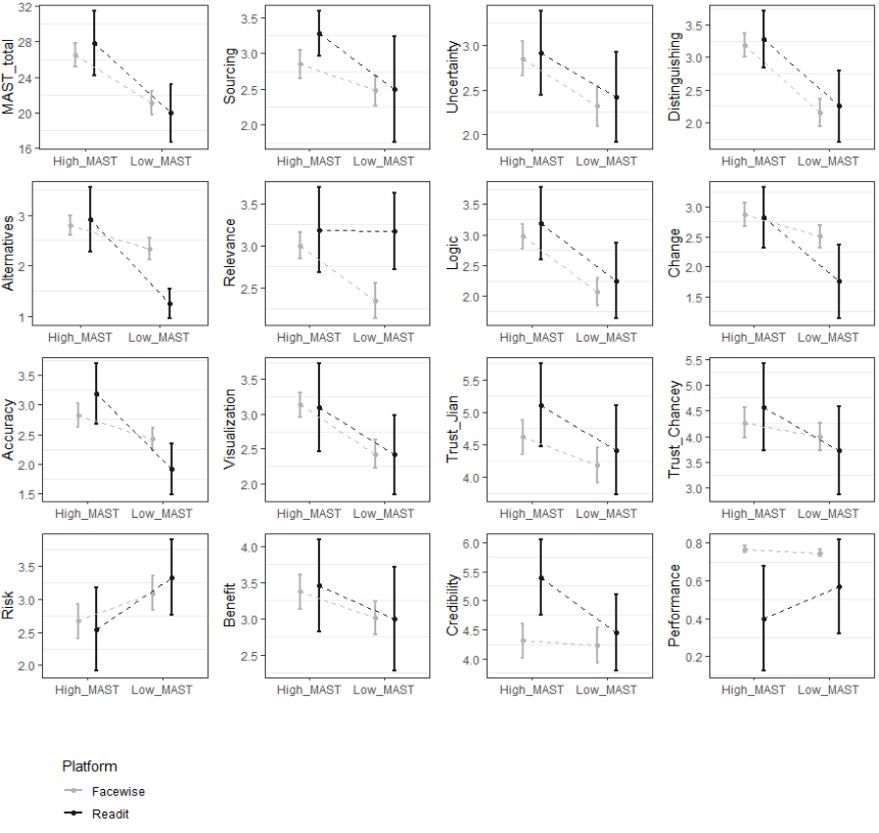}
    \caption{Means with 95\% Confidence Intervals for Facewise and READIT across different levels of Low-MAST and High-MAST. We used \includegraphics[height=2ex]{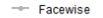}  for Facewise and \includegraphics[height=2ex]{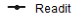} for READIT.}
    \label{fig:figF1}
\end{figure}

\newpage 
\section*{Appendix F: Participant demographics for Facewise and READIT}
\label{appendixF}

\begin{table*}[h]
\small
  \begin{center}
    \begin{tabular}{p{0.35\linewidth}|p{0.28\linewidth}|p{0.28\linewidth}}
      \toprule 
      & \textbf{High-MAST} (\textit{n} = 73) & \textbf{Low-MAST} (\textit{n} = 73) \\ 
      \midrule 
      \multirow{2}{*}{\textbf{Years of experience as a TSO}} & 55\% 3 years or less & 50\% 3 years or less \\
                                                    & 24\% 10 or more years	& 28\% 10 or more years \\

      \midrule 
      \multirow{2}{*}{\textbf{Highest degree}}  & 69\% 2-year college or less & 74\% 2-year college or less  \\
                                                & 26\% 4-year college & 21\% 4-year college \\
      \midrule 
      \textbf{Volunteer hours in the past 3 months}  & 62\% 0 hours & 71\% 0 hours \\
      \midrule 
      \textbf{Computer habit} & 58\% daily & 65\% daily \\
      \midrule 
      \multirow{2}{*}{\textbf{Gaming habit}}    & 18\% daily 26\% never & 30\% daily\\
                                                & 26\% never & 17\% never \\
      \midrule 
      \multirow{2}{*}{\textbf{Screen hours before study}}   & Mean: 2 hrs.  & Mean: 2.2 hrs.\\
                                                            & Median: 1.2 hrs. & Median: 2 hrs. \\
      \bottomrule 
    \end{tabular}
    \caption{Participant demographics across High-MAST and Low-MAST for Facewise.}
    \label{tab:table1}
  \end{center}
\end{table*}

\begin{table}[h]
\small
  \begin{center}
    \begin{tabular}{p{0.37\linewidth}|p{0.27\linewidth}|p{0.27\linewidth}}
      \toprule 
      & \textbf{High-MAST} (\textit{n} = 11) & \textbf{Low-MAST} (\textit{n} = 12) \\ 
      \midrule 
      \multirow{3}{*}{\textbf{Age}} & 36\% 30 years or less   & 34\% 30 years or less \\
                                    & 18\% 31-39 years       & 33\% 31-39 years\\
                                    & 46\% 40 or more years   & 33\% 40 or more years\\
      \midrule 
      \multirow{2}{*}{\textbf{Gender}}  & 73\% man & 50\% man  \\
                                        & 27\% woman & 50\% woman \\
      \midrule 
      \textbf{Race}  & 82\% white & 83\% white \\
      \midrule 
      \multirow{3}{*}{\textbf{Years of experience as an IA}}    & 18\% 2 years or less  & 25\% 2 years or less\\
                                                                & 27\% 3-5 years        & 17\% 3-5 years\\
                                                                & 55\% 6 years or more  & 58\% 6 years or more\\
      \midrule 
      \textbf{Experience with AI-DSS} & 46\% no prior experience & 33\% no prior experience \\
      \midrule 
      \multirow{3}{*}{\textbf{Highest degree}}  & 27\% 4-year college   & 17\% 4-year college\\
                                                & 73\% master’s         & 66\% master’s \\
                                                &                       & 17\% doctorate \\
      \midrule 
      \textbf{Experience with VAST challenge} & 100\% no & 100\% no\\
      \midrule 
      \textbf{Experience with clustering tools} & 55\% no & 33\% no \\
      \midrule 
      \multirow{2}{*}{\textbf{Screen hours before study}}   & Mean: 5.5 hrs.& Mean: 5 hrs.\\
                                                            & Median: 6 hrs. & Median: 5 hrs.\\

    \bottomrule 
    \end{tabular}
    \caption{Participant demographics across High-MAST and Low-MAST for READIT.}
  \label{tab:table2}
  \end{center}
\end{table}

\vskip 0.2in

\end{document}